\documentclass[a4paper,fleqn,usenatbib]{mnras}

\usepackage[T1]{fontenc}
\usepackage{ae,aecompl}

\usepackage{ulem}

\usepackage{graphicx}	
\usepackage{amsmath}	
\usepackage{amssymb}	

\newcommand{\Myr}{\,\mathrm{Myr}}
\newcommand{\Gyr}{\,\mathrm{Gyr}}

\newcommand{\feh}{[{\rm Fe}/{\rm H}]}
\newcommand{\afe}{[\alpha/{\rm Fe}]}
\newcommand{\xfe}{[{\rm X}/{\rm Fe}]}
\newcommand{\mgfe}{[{\rm Mg}/{\rm Fe}]}
\newcommand{\ofe}{[{\rm O}/{\rm Fe}]}
\newcommand{\sife}{[{\rm Si}/{\rm Fe}]}
\newcommand{\cafe}{[{\rm Ca}/{\rm Fe}]}
\newcommand{\kpc}{\,{\rm kpc}}

\newcommand{\logg}{\log(g)}
\newcommand{\Teff}{T_{\rm eff}}

\usepackage{natbib}
\usepackage{mathtools}

\usepackage[dvipsnames]{xcolor}

\def\gtsim {>\kern-1.2em\lower1.1ex\hbox{$\sim$}~}   
\def\ltsim {<\kern-1.2em\lower1.1ex\hbox{$\sim$}~}   

\usepackage{newtxtext,newtxmath}

\title[Distribution of $\alpha$/Fe in the MW disc]{The distribution of [$\alpha$/Fe] in the Milky Way disc}

\author[F. Vincenzo et al.]{Fiorenzo Vincenzo$^{1}$\thanks{email: vincenzo.3@osu.edu}, David H. Weinberg$^{1,2}$\thanks{email: weinberg.21@osu.edu}, Andrea Miglio$^{3}$, Richard R. Lane$^{4}$, 
\newauthor Alexandre Roman-Lopes$^{5}$
\\ ~ \\
$^{1}$Department of Astronomy \& Center for Cosmology and AstroParticle Physics, The Ohio State University, Columbus, OH 43210, USA  \\
$^{2}$Institute for Advanced Study, Princeton, NJ 08540, USA \\
$^{3}$School of Physics and Astronomy, University of Birmingham, Edgbaston, B15 2TT, UK \\
$^{4}$Instituto de Astronomía y Ciencias Planetarias de Atacama, Universidad de Atacama, Copayapu 485, Copiapó, Chile \\
$^{5}$Departamento de Física, Facultad de Ciencias, Universidad de La Serena, Cisternas 1200, La Serena, Chile
}

\begin{document}

\date{Accepted 2021 October 03. Received 2021 October 03; in original form 2021 January 12 }

\pagerange{\pageref{firstpage}--\pageref{lastpage}} \pubyear{2021}

\maketitle

\label{firstpage}


\begin{abstract}
Using a sample of red giant stars from the Apache Point Observatory Galactic Evolution Experiment (APOGEE) Data Release 16, we infer the conditional distribution $p(\afe\,|\,\feh)$ in the Milky Way disk for the $\alpha$-elements Mg, O, Si, S, and Ca.  In each bin of $\feh$ and Galactocentric radius $R$, we model $p(\afe)$ as a sum of two Gaussians, representing ``low-$\alpha$'' and ``high-$\alpha$'' populations with scale heights $z_1=0.45\kpc$ and
$z_2=0.95\kpc$, respectively.  By accounting for age-dependent and $z$-dependent selection effects in APOGEE, we infer the $\afe$ distributions that would be found for a fair sample of long-lived stars covering all $z$. Near the Solar circle, this distribution is bimodal at sub-solar $\feh$, with the low-$\alpha$ and high-$\alpha$ peaks clearly separated by a minimum at intermediate $\afe$.  In agreement with previous results, we find that the high-$\alpha$ population is more prominent at smaller $R$, lower $\feh$, and larger $|z|$, and that the sequence separation is smaller for Si and Ca than for Mg, O, and S. We find significant intrinsic scatter in $\afe$ at fixed $\feh$ for both the low-$\alpha$ and high-$\alpha$ populations, typically $\sim 0.04$-dex.  The means, dispersions, and relative amplitudes of this two-Gaussian description, and the dependence of these parameters on $R$, $\feh$, and $\alpha$-element, provide a quantitative target for chemical evolution models and a test for hydrodynamic simulations of disk galaxy formation.  We argue that explaining
the observed bimodality will probably require one or more sharp transitions in the disk's gas accretion, star formation, or outflow history in addition to radial mixing of stellar populations.
\end{abstract}


\begin{keywords}
Galaxy: abundances -- Galaxy: disc -- Galaxy: stellar content -- stars: abundances 
\end{keywords}

 
\section{Introduction} \label{sec:intro}

The distribution of stars in the space of $\alpha$-element abundances and
iron abundances is a powerful diagnostic of the star formation history (SFH)
and stellar nucleosynthesis in galaxies, as the $\alpha$-elements come 
primarily from core collapse supernovae (CCSN) while iron comes also from
Type Ia supernova (SNIa) enrichment with a longer timescale
(e.g., \citealt{tinsley1980,matteucci1986,mcwilliam1997}).
The evolutionary track of a stellar population in $\afe-\feh$ provides
a diagnostic of star formation efficiency, accretion history, and gas
outflows, as well as supernova yields 
(see, e.g., \citealt{matteucci2012,andrews2017,weinberg2017}, 
hereafter WAF).\footnote{We follow standard abundance notation in which
[X/Y] = $\log X/X_\odot - \log Y/Y_\odot$.}
Stars with kinematics or geometry characteristic of the ``thick disk''
\citep{gilmore1983} exhibit higher values of $\afe$ at a given $\feh$
\citep{fuhrmann1998,bensby2003}.  This separation is clear enough that
many studies now use $\afe$ ratios in place of kinematics to define a
thick disk population (e.g., \citealt{lee2011,haywood2013,mackereth2017}).
However, the correlation of abundance patterns with geometry, kinematics,
and stellar age makes the distribution of stars in $\afe-\feh$ highly
dependent on sample selection, complicating the challenge of using this
distribution to infer the enrichment history of the Milky Way.

In this paper we examine the $\afe-\feh$ distribution of disk stars in
the Apache Point Observatory Galactic Evolution Experiment
(APOGEE; \citealt{majewski2017}).  In some zones of Galactocentric radius $R$
and midplane distance $|z|$, this distribution appears distinctly bimodal,
with two separated tracks at low $\feh$ that merge at $\feh \ga 0$
(\citealt{hayden2015}, hereafter H15).
However, in other zones the bimodality is less clear, and it is not
obviously present in some other studies such as \citet[fig.~15]{edvardsson1993} or Data Release 2 (DR2) of the
Galactic Archaeology with HERMES (GALAH) survey
(\citealt{buder2018}, fig.~22; but see fig.~5 of the DR3 paper
\citealt{buder2020}).
The sample selection may differ a lot in the various surveys, as -- for example -- APOGEE focuses on red giants, whereas GALAH focuses on dwarfs. Dwarfs and giants have their own issues with respect to abundance determinations. In this context, an interesting spectroscopic survey is the RAdial Velocity Experiment (RAVE; e.g., see fig.~22 in \citealt{guiglion2020}), which probes a composite population of dwarfs and red giants. 

For our analysis here, we fit an empirical model to the APOGEE red giant
disk population, accounting for selection effects, to infer the
{\it intrinsic} distribution of $\afe$ ratios in bins of $\feh$ and $R$
for the $\alpha$-elements Mg, O, Si, S, and Ca.
By intrinsic, we mean the distribution that would be measured if one
could observe an unbiased subset of all disk stars with main sequence
lifetimes longer than the age of the Galaxy.  This intrinsic distribution
can be compared to predictions of numerical simulations or 
galactic chemical evolution (GCE) calculations without modeling the
detailed geometry and sample selection of APOGEE.

A variety of scenarios have been proposed to explain the co-existence of
high-$\alpha$ and low-$\alpha$ populations.\footnote{Arguably ``low-Ia''
and ``high-Ia'' is a better nomenclature \citep{Griffith2019}, since the physical difference
between these populations is the amount of Fe-peak enrichment from SNIa.
In this paper we will stick with the conventional empirical nomenclature
based on the degree of ($\alpha$/Fe) enhancement relative to the solar
ratio.}
The two-infall model posits two waves of star formation separated by a period
of pristine gas accretion with minimal star formation. The first wave 
produces the high-$\alpha$ population, and infall resets the gas phase
metallicity to low values before the second wave produces the low-$\alpha$
population (e.g., see \citealt{chiappini1997,noguchi2018,spitoni2019,palla2020}, but also \citealt{lian2020}). In this context, \citet{vincenzo2019} proposed that the physical mechanism responsible for a temporary quenching of the star-formation history of the MW at high redshifts was a major merger event with a galaxy like Gaia-Enceladus or Gaia-Sausage \citep{helmi2018,belokurov2018}; this merger may have also kinematically heated part of the present-day inner halo and thick-disk (e.g., \citealt{haywood2018,chaplin2020,montalban2020}). A possible scenario could be that the merger event between the MW progenitor and Gaia-Sausage/Enceladus activated the central black hole of our Galaxy at high redshifts, heating up the gas in the DM halo over a limited period of time, eventually giving rise to a second gas accretion event which brought the MW back to the main sequence of star-forming disk galaxies (see the simulations of \citealt{pontzen2017} for the feasibility of this scenario). 

In contrast to the two-infall picture,
\cite{schoenrich2009} argue that the low-$\alpha$ population is not an
evolutionary sequence at all but instead comprises the end-points of
evolutionary tracks at different radii, which are then mixed by
stellar migration 
(see also \citealt{nidever2014,sharma2020}). 
Conversely, \cite{clarke2019} suggest that the low-$\alpha$ population is the true
evolutionary sequence and the high-$\alpha$ population forms in massive
clumps of the gas rich early disk, which self-enrich with CCSN elements. 
A similar phenomenon is seen in the simulations of \cite{vincenzo2020} and
\cite{khoperskov2020}. 

The success of hydrodynamic cosmological simulations in producing distinct $\afe-\feh$ sequences is mixed. In these simulations, stellar migrations, star-formation and chemical enrichment, outflows and inflows of gas, and cosmological growth are all self-consistently taken into account. Analyzing the EAGLE cosmological volume and the Auriga zoom-in simulations, respectively, \cite{mackereth2018} and \cite{grand2018} find that bimodality of $\afe$ arises in a small fraction ($\sim 10\%$) of simulated galaxies that have uncommon, 2-phase accretion histories.  However, \cite{buck2020} finds bimodality in all four simulations analyzed from the NIHAO-UHD suite, a consequence of gas rich mergers that bring in low metallicity fuel, reminiscent of the 2-infall picture.  An interesting simulation-based study is that of  \cite{vincenzo2020}, who can qualitatively reproduce both the bimodality in [$\alpha$/Fe]-[Fe/H] and the slope of the metallicity gradient as observed by APOGEE-DR16. In this simulation, accretion and gas flows play the primary role in determining [$\alpha$/Fe]-[Fe/H]-age distributions, and stellar migration has a secondary impact.

Testing the predictions of models and simulations requires an accurate
quantification of the intrinsic $\afe$ distribution across the
Milky Way (MW) disk.  There are two reasons one cannot simply use the observed
$\afe$ ratios of the APOGEE sample to provide this distribution.
The first is that higher $\afe$ stars are systematically older, and 
the fraction of a stellar population's original stars that are red
giants at a given time depends on the population's age.  
The second is that higher $\afe$ stars have a larger scale height, and
the APOGEE sample has a complex selection in $|z|$
(\citealt{zasowski2013,zasowski2017}; see fig.~\ref{fig:z-correction} below).
For each $\alpha$-element that we consider, we model the conditional 
distribution $p(\afe)$ in bins of $R$ and $\feh$ as a double-Gaussian,
parameterized by the means and dispersions of the high-$\afe$ and low-$\afe$
populations and by their relative normalization, adopting empirically
motivated scale heights for each population.
Our analysis covers the range $3\kpc \leq R \leq 11\kpc$,
$0 \leq |z| \leq 2\kpc$, and $-0.5 \leq \feh < 0.2$.

We adopt a two-component form because it provides a good description of 
the data, but we make no assertion about whether these components have
physically distinct origins.  In the case of the vertical distribution 
of disk stars, \cite{bovy2012} argue that the double-exponential form
suggesting distinct thin and thick disks emerges from a continuous
dependence of single-exponential scale heights on abundances, which
are themselves correlated with population age.
While the degree of bimodality in $\afe$ is our primary target, the
intrinsic scatter within each population is also a quantity of interest,
an important test of stochastic enrichment models and radial mixing
of populations.  The relative mean sequences for the different elements
constrains the relative contribution of SNIa enrichment to that element
(WAF; \citealt{weinberg2019}).

Our work is organised as follows. In Section \ref{sec:data}, we describe the assumed data set and discuss the main selection effects. In Section \ref{section:model} we describe the formalism and the assumptions of our model and present the methods employed to fit the observed chemical abundance distributions. In Section \ref{sec:results}, we present our findings for the conditional distributions of $p([\alpha/\text{Fe}])$ in bins of $R$ and $\text{[Fe/H]}$, including the degree of bimodality, the intrinsic scatter for high-$\alpha$ and low-$\alpha$ populations, and the dependence on the choice of $\alpha$-elements. We summarize our results and discuss implications in Section \ref{sec:conclusions}.

\begin{figure*}
\centering
\includegraphics[width=17cm]{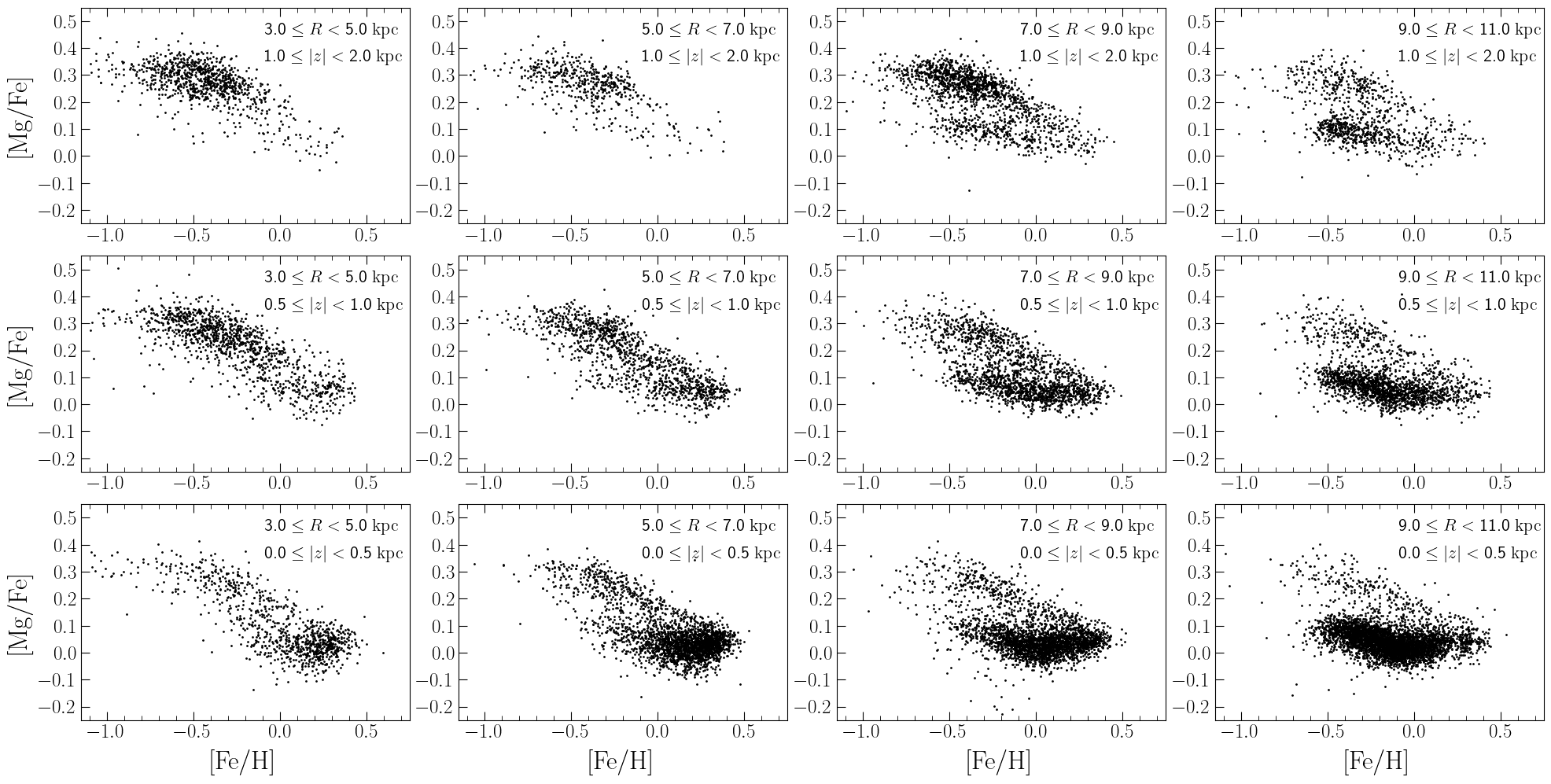}
\caption{The observed [Mg/Fe] versus [Fe/H] abundance diagram for the adopted sample of stars from SDSS-IV APOGEE-2 DR16 \citep{ahumada2020}. Different columns correspond to different ranges of Galactocentric distance, $R$, whereas different rows correspond to different ranges in height above/below the Galactic plane, $z$.  }
\label{fig:alphafe-feh-data}
\end{figure*}

\begin{figure}
\centering
\includegraphics[width=6.5cm]{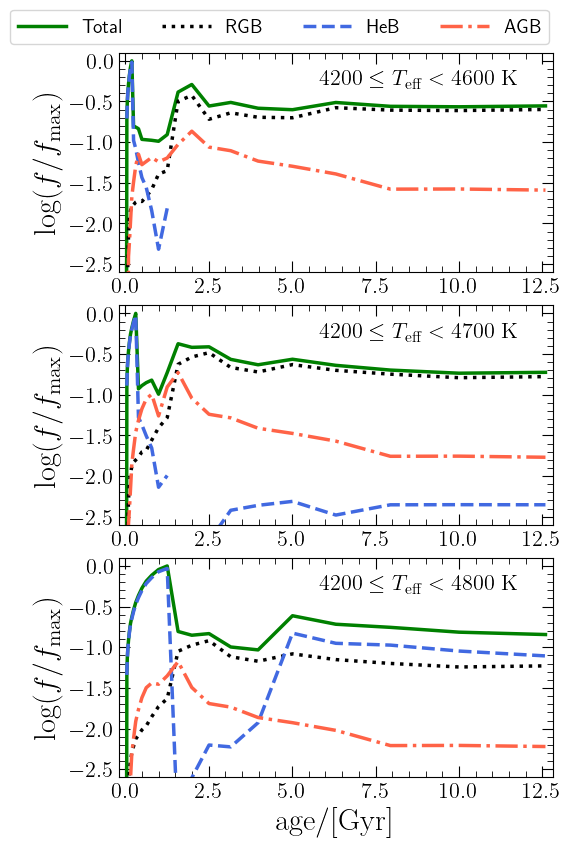}
\caption{The selection factor $f$ as derived from the \textsc{PARSEC} stellar evolutionary tracks (release v1.2S) with metallicity $Z=0.01$ \citep{bressan2012,tang2014,chen2015}, by assuming the IMF of \citet{kroupa2001}. Different panels correspond to different maximum values of $T_{\text{eff}}$, from $4600$ to $4800\;\text{K}$. }
\label{fig:correction-factor}
\end{figure}

\begin{figure}
\centering
\includegraphics[width=6.5cm]{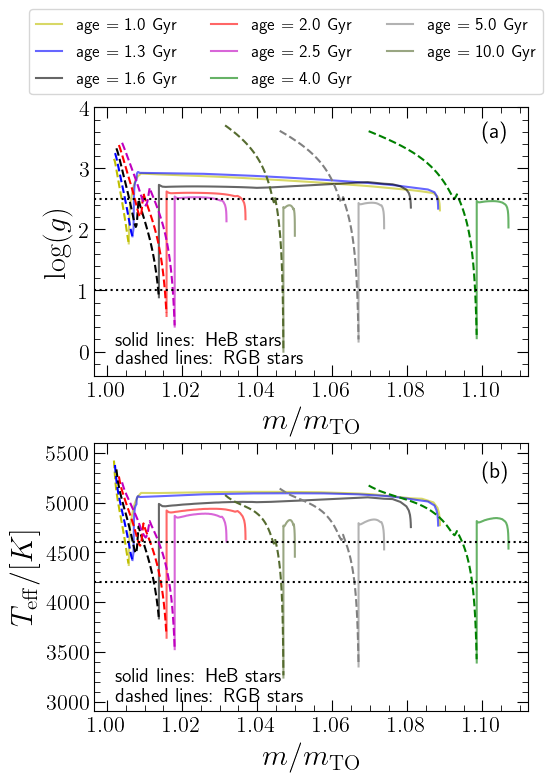}
\caption{\textit{(a)} The evolution of $\log(g)$ as a function of the stellar mass, as normalised with respect to the turn-off mass (maximum mass on the main sequence at the given age). Dashed lines correspond to red giant stars, whereas solid lines correspond to He-burning stars. Different colours correspond to different ages of the stars. Horizontal dotted lines mark the assumed selection window. \textit{(b)} Same as in the previous panel, but showing the evolution in the effective temperature, $T_{\text{eff}}$.  Although evolution in $m$ is monotonic with age, the evolution in $m/m_{\text{TO}}$ is not: curves for ages 1-2.5 Gyr are ordered left to right, while the curves for 10, 5, and 4 Gyr are ordered right to left.
}
\label{fig:selection-characterisation}
\end{figure}

\section{APOGEE data and Selection Effects} 
\label{sec:data}

\subsection{Data Sample}

Our sample is selected from SDSS-IV APOGEE-2 DR16 \citep{ahumada2020} with the following cuts in effective temperature, surface gravity, and signal-to-noise ratio (SNR):
\begin{enumerate}
    \item Effective temperature in the range $4200 \le T_{\text{eff}} < 4600\;\text{K}$, 
    \item Surface gravity in the range $1 \le \log(g/[\text{cm}\,\text{s}^{-2}]) <2.5$,
    \item Signal-to-noise ratio $\text{SNR} > 80$.  
\end{enumerate}
The abundance determination methods and calibrations are described by \cite{holtzman2015}, \cite{garcia-perez2016}, and \cite{jonsson2020}.
In Fig. \ref{fig:alphafe-feh-data} we show how the observed [Mg/Fe]-[Fe/H] abundance diagram varies within the Galactic disc, for different ranges of Galactocentric radius $R$ and height $|z|$, given the adopted selections. The distances are derived with \textsc{AstroNN} \citep{leung2019a,leung2019b}\footnote{\textsc{AstroNN}'s documentation can be visited at the following link: \url{https://astronn.readthedocs.io/en/latest/}.} and are publicly available as a value-added catalogue for APOGEE-DR16\footnote{The APOGEE-DR16 value-added catalogue with distances from \textsc{AstroNN} can be downloaded at the following link: \url{https://www.sdss.org/dr16/data_access/value-added-catalogs/?vac_id=the-astronn-catalog-of-abundances,-distances,-and-ages-for-apogee-dr16-stars}.}. Our analysis focuses on stars with iron abundances in the range $-0.5\le \text{[Fe/H]} \le 0.2$.

\subsection{Age-Selection Effects}
\label{sec:age-selection}

Selection effects may arise because we limit our analysis to stars in a specific range of $\log(g)$ and $T_{\text{eff}}$. This selection can be age-dependent because the fraction of stars within the selection window changes with the age of the population.  We account for this effect by including in our model a selection factor, which is computed as  
\begin{equation}
\label{eq:corr}
    f(\tau) = \frac{ \int_{M_{a}(\tau)}^{M_{b}(\tau)}{ dM\,\frac{dN}{dM} } }{ \int_{M_{\text{min}}}^{M_{\text{max}}(\tau)}{ dM\,\frac{dN}{dM} } }, 
\end{equation}
where $dN/dM$ is the initial mass function (IMF) and $M_{a}$ and $M_{b}$ are the initial masses of the stars that are in the $\log(g)$ and $T_{\text{eff}}$ range for a population of age $\tau$.  We assume that all stars in this range have an equal chance of being selected as APOGEE targets; the impact of this assumption on the age distribution of the stars is quantified in Section \ref{sec:targetsel}. Finally, the quantities $M_{\text{min}}$ and $M_{\text{max}}$ in equation (\ref{eq:corr}) denote the current mass range of the IMF, but the denominator cancels out in our analysis so the choice is arbitrary. It is the \textit{number} fraction that matters, not the mass fraction, because we are counting stars. Because the observed age-$\afe$ relation is much tighter than the observed age-$\feh$ relation, we will use $\afe$ to compute $f(\tau)$ and ignore any dependence on $\feh$ at fixed $\afe$.

In Fig. \ref{fig:correction-factor} we show how $f$ varies as a function of the stellar age, by adopting the \textsc{PARSEC} stellar evolutionary tracks (release v1.2S) at $Z=0.01$ \citep{bressan2012,tang2014,chen2015} and the IMF of \citet{kroupa2001}. Different panels correspond to different maximum values of $T_{\text{eff}}$ in the selection window, from $T_{\text{max}} = 4600\;\text{K}$ to $4800\;\text{K}$, since $f$ does not change considerably for maximum temperatures $T_{\text{eff,max}}>4800\;\text{K}$. By decreasing $T_{\text{eff,max}}$ from $4800\;\text{K}$ to $4600\;\text{K}$, we can effectively suppress the contribution of stars in the red clump (which dominate at ages $\gtrsim 4\;\text{Gyr}$) and in the secondary clump (which dominate at ages $\lesssim 2\;\text{Gyr}$; we address the readers to \citealt{girardi1999,girardi2016} for details). In Fig. \ref{fig:selection-characterisation} we show how red-giant branch and He-burning stars of different age evolve in $\log(g)$ and $T_{\text{eff}}$ as a function of the stellar mass, normalized to the main-sequence turn-off mass (maximum mass on the main-sequence for each stellar age in the figure), by considering isochrones of different age. The secondary clump in Fig. \ref{fig:selection-characterisation} is represented by the extended He-burning tracks at ages $<2\;\text{Gyr}$.

The encouraging feature of Fig.~\ref{fig:correction-factor} is that, for $T_{\text{max}} = 4600\;{\rm K}$, the selection factor $f$ is nearly constant for $\tau > 2.5\Gyr$.  There is a moderate rise at $\tau \approx 2\Gyr$, where a somewhat larger initial mass range resides in the $\log(g)-T_{\text{eff}}$ range (see Fig.~\ref{fig:selection-characterisation}).  There is a more pronounced deficit at $\tau \approx 1\Gyr$, where tracks do not fully cross $\log(g)=1-2.5$.  
There is a spike at $\tau\approx 100\Myr$, but this is very narrow.  The approximate constancy of $f$ means that moderate errors in typical stellar ages will have little impact on the recovered $\afe$ distributions.  Ranges of $\afe$ that are sharply peaked near ages of $1\Gyr$ or $2\Gyr$ will have larger corrections and uncertainties.

\begin{figure}
\centering
\includegraphics[width=6.5cm]{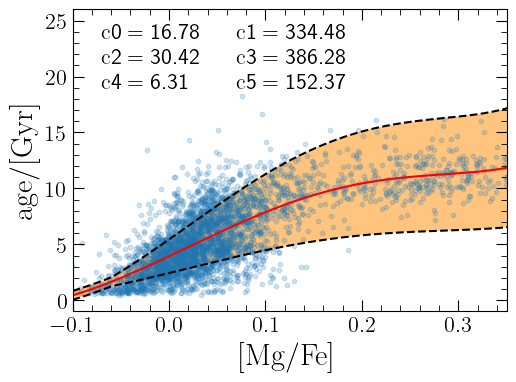}
\caption{The age-[Mg/Fe] relation assumed in our model, in order to compute the selection factor shown in Fig. \ref{fig:correction-factor}. The blue points represent the sample of \citet{miglio2020}, and the red curve corresponds to the best fit with a series of Legendre polynomials of degree $5$ (the coefficients for the best fit are reported in the figure). The orange area shows the asteroseismic errors as a function of the stellar age, from the analysis of \citet{miglio2020}.   }
\label{fig:age-mgfe}
\end{figure}

When we fit our model to the observed $[\alpha/\text{Fe}]$ distributions in the Galaxy disc, we need an estimate of the age of the stars to compute the correction factor, $f$. In this work, we fit the age-[Mg/Fe] relation as measured by the asteroseismic analysis of \citet[see Fig. \ref{fig:age-mgfe}]{miglio2020} for a sample of stars in the Kepler fields. For the fitting function, we adopt a series of Legendre polynomials of degree $5$. The results of our analysis are shown in Fig. \ref{fig:age-mgfe}, where the coefficients of the best fit are also reported. 

We include dispersion of the age-[Mg/Fe] relation in our calculations by fitting the asteroseismic measurement errors as a function of the stellar age. The best-fit description has the following functional form: 
\begin{equation} \label{eq:age-alpha-miglio}
    \sigma_{\text{age}} = \begin{cases}  0.482\times\frac{\text{age}}{\text{Gyr}} -0.395\,\text{\text{Gyr}} & \text{if}\;\text{age} > 1.65\,\text{Gyr} \\
    0.4\;\text{Gyr} &  \text{if}\;\text{age}\le 1.65 \, \text{Gyr} \end{cases}
\end{equation}
We enforce a minimum dispersion of $0.4\;\text{Gyr}$. Equation (\ref{eq:age-alpha-miglio}) represents observational errors rather than intrinsic dispersion, but the dispersion plays only a small role in our analysis, so this rough estimate is sufficient. In Fig. \ref{fig:age-mgfe}, stars with $\mgfe > 0.1$ lie well within the orange band implied by equation (\ref{eq:age-alpha-miglio}) because the age error is predominantly systematic rather than a random error that adds star-by-star scatter.

Even though the sample of \citet{miglio2020} mostly comprises stars in the Solar neighborhood, we apply equation (\ref{eq:age-alpha-miglio}) to the inner and outer annuli, being aware that the [Mg/Fe]-age relation may change as a function of the Galactocentric distance, as predicted by chemical evolution models and simulations (e.g., see \citealt{vincenzo2020}).

\subsection{$z$-Selection Effects}
\label{sec:z-selection}
 Our model fits the observed distribution of stars in the ([Mg/Fe], $|z|$) plane, for different bins of $R$ and [Fe/H]. However, APOGEE does not sample correctly the true vertical distribution of the stars in the MW disc, since the observations are limited by the number of fibers on the plates in the focal plane of the telescope and by the sky coverage of the survey. For example, at many Galactic longitudes APOGEE observed five fields at $b=-8$, $-4$, 0, $+4$, and $+8$, obtaining spectra for $\sim 250$ science targets in each 7-deg$^2$ field.  In the high density midplane these 250 targets are only a small fraction of the available giants that pass the APOGEE magnitude cut, while at higher latitudes the fraction is larger.  Thus, this observing strategy tends to overrepresent high-$|z|$ stars in the sample.

\begin{figure}    
\centering
\includegraphics[width=6.5cm]{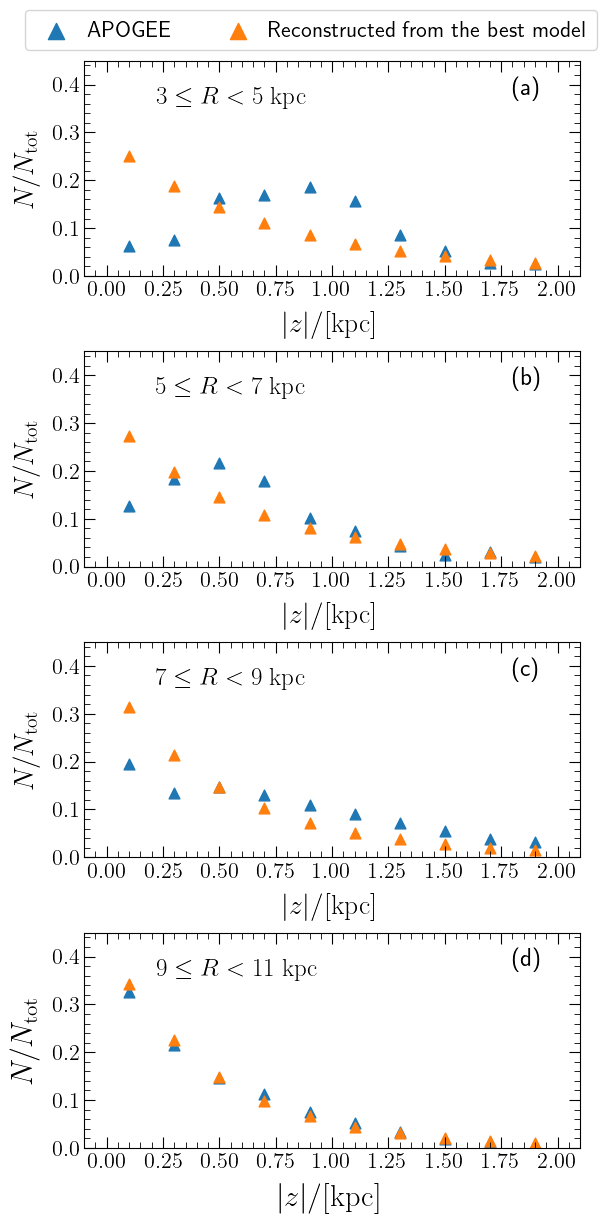}
\caption{The normalized number of stars in our APOGEE-DR16 sample (blue triangles) as a function of $|z|$, for different ranges of Galactocentric distance, $R$ (different panels). The orange triangles correspond to the predictions of our best model in which $|z|$-selection effects are removed, assuming that APOGEE correctly samples the MDF in $-0.4 \le \text{[Fe/H]} < 0$. }
\label{fig:z-correction}
\end{figure}

The importance of assessing the $z$-selection effects is demonstrated in Fig. \ref{fig:z-correction}, in which we compare the vertical distribution of stars as observed by APOGEE-DR16 (for our adopted data cuts) with the predictions of our best model, using uniform bins in $|z|$ of $0.2\;\text{kpc}$.
Since the parameters of the best model depend on [Fe/H], in order to make Fig. \ref{fig:z-correction} we assume that APOGEE correctly samples the metallicity distribution function (MDF) in the range $-0.4 \leq \mathrm{[Fe/H]}  <  0.0$, and we compute the re-scaling factors for the model predictions to match the integrated number of stars over $|z|$ as observed by APOGEE in the considered range of [Fe/H].
Fig.~\ref{fig:z-correction} shows that in the inner Galaxy APOGEE overrepresents stars with $|z|\approx 1\kpc$ and underrepresents stars near the midplane.  This effect gradually weakens with increasing $R$, and for $9 \leq R < 11\kpc$ the observed $|z|$ distribution coincides well with the true $|z|$ distribution predicted by our best-fit model.

\begin{figure*}    
\centering
\includegraphics[width=17cm]{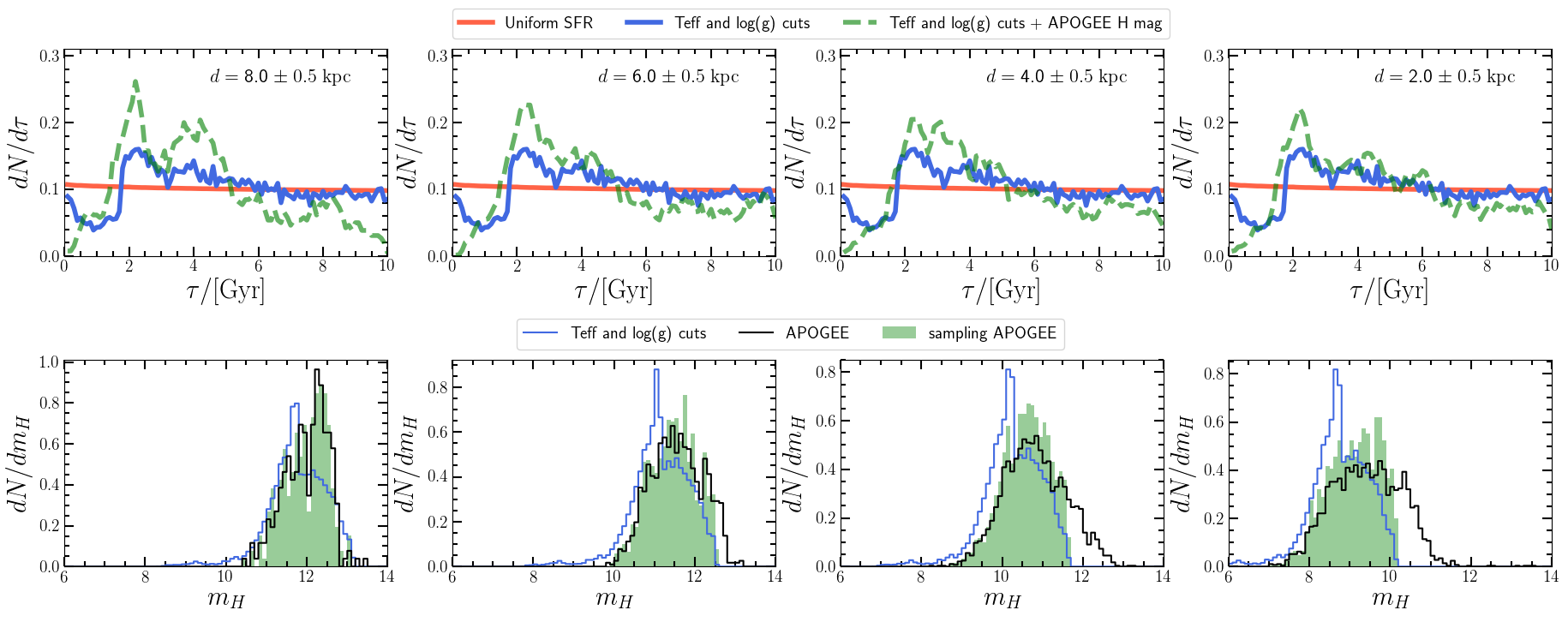}
\caption{Impact of apparent-magnitude selection on the sample age distribution.  In the upper panels, red curves (same in each panel) show the age distribution $dN/d\tau$ of surviving stars formed with a constant SFR.  Blue curves (again the same in each panel) show the age distribution of stars satisfying our sample selection criteria,
$1 \le \log(g) < 2.5$ and $4200 \le T_{\mathrm{eff}} < 4600\,\text{K}$, obtained by randomly sampling the IMF with $10^6$ stars from 100 isochrones separated at 0.1-Gyr intervals.  In the lower panels, black histograms show the $m_{H}$ distribution of stars in our APOGEE sample in heliocentric distant ranges centred on $d=8$, 6, 4, and 2 kpc (left to right).  Blue histograms show the predicted distributions of the Monte Carlo red giant sample, which peak at brighter apparent magnitudes.  By weighted resampling of this sample we obtain the green shaded histograms and the age distributions shown by the green-dashed curves in the upper panels; these have been convolved with a 1 Gyr rectangular window to reduce noise.  Apparent magnitude selection has only moderate impact on the age distribution beyond that already arising from $\logg$ and $\Teff$ selection.
}
\label{fig:mh-selection}
\end{figure*}

\subsection{Apparent Magnitude Selection} \label{sec:targetsel}

The selection of APOGEE stars in apparent $H$-band magnitude $m_H$ is fairly complex, designed to produce a sample that maps the Galaxy effectively with an efficient observing strategy \citep{zasowski2013,zasowski2017}.  This apparent magnitude selection can affect the relative probability of observing stellar populations of different ages, beyond the age-selection imposed by the $\logg$ and $\Teff$ cuts which we have accounted for as described in \S\ref{sec:age-selection}.  To test the possible importance of this effect, we have created Monte Carlo samples using PARSEC isochrones (see \S\ref{sec:age-selection}) spaced at 0.1-Gyr intervals from 0.1-10 Gyr.  For each isochrone we make $10^6$ random draws from the \cite{kroupa2001} IMF and compute the $\logg$, $\Teff$, and absolute $H$-band magnitude $M_H$ of the resulting stars, dropping any that have evolved past the AGB.  We can then apply $\logg$, $\Teff$, and distance-dependent $M_H$ cuts to see how they affect the age distribution $dN/d\tau$.

In the upper panels of Fig.~\ref{fig:mh-selection}, red curves (the same in each panel) show the age distribution of all surviving stars.  Because we form equal numbers of stars in each time interval, these curves are nearly flat, declining slightly towards larger ages because a larger fraction of the IMF has evolved past the AGB.  Blue curves (again the same in each panel) show the effect of imposing our $4200\;{\rm K} \leq \Teff < 4600\;{\rm K}$ and $1 \leq \logg < 2.5$ cuts.  As shown previously in Fig.~\ref{fig:correction-factor}, populations of age $\tau \approx 2\Gyr$ are moderately overrepresented while populations younger than 2 Gyr are underrepresented.

In the lower panels of Fig.~\ref{fig:mh-selection}, black histograms show the $m_H$ distribution of stars in our APOGEE sample in four ranges of distance from the sun, centred at $d=2$, 4, 6, and 8 kpc.  For these histograms we have selected stars with $|z| \leq 0.5\kpc$.  Blue histograms show the distribution of $m_H = M_H + \mu(d)$ for stars in our Monte Carlo sample satisfying our $\logg$ and $\Teff$ selection cuts, where $\mu(d)$ is the distance modulus at the bin center.  At small distances these histograms are peaked at lower $m_H$ than the observed sample distribution.  Finally, we draw stars from this Monte Carlo sample with weighting in $M_H$ to reproduce, approximately, the observed $m_H$ distributions, obtaining the green shaded histograms.  The observed histograms extend to fainter $m_H$ at small distances, which simply indicates that stars with $\logg < 2.5$ are too luminous to have a large $m_H$ at small $d$.

Returning to the upper panels, green curves show $dN/d\tau$ for this reweighted sample.  These curves are noisy because the number of stars passing all cuts for a given isochrone is small, and we have smoothed them with a $1\Gyr$ boxcar filter to improve readability.  For the most part, the magnitude selection produces only moderate changes to $dN/d\tau$ beyond those already caused by the $\logg$ and $\Teff$ selection.  The most significant differences are a somewhat higher overrepresentation of 2-4 Gyr old stars and a suppression of very young populations with $\tau < 0.5\Gyr$, whose $\logg < 2.5$ giants are sufficiently luminous that they lie off the bright end of the APOGEE $m_H$ histograms.

In principle we could incorporate $m_H$ selection effects in our modeling by fitting the observed counts in narrow bins of $\afe$, $|z|$, {\it and} $m_H$, within each larger bin of [Fe/H] and $R$.  Because the effects in Fig.~\ref{fig:mh-selection} are fairly small, we have chosen not to incur this additional complication, making only the corrections described in \S\ref{sec:age-selection} and \S\ref{sec:z-selection}.  Somewhat fortuitously, the age-selection effects in APOGEE are nearly flat except for narrow ranges of $\tau$, and the $|z|$-selection effects illustrated in Fig.~\ref{fig:z-correction} have greater impact.  As a further test, we have checked that the $M_H$ distributions of the observed low-$\alpha$ and high-$\alpha$ stars in a given bin of $R$, $|z|$, and $\feh$ largely overlap.  For $R > 7\kpc$ the low-$\alpha$ stars, which are younger on average, are shifted towards brighter $M_H$ by 0.5-1 mag, but this difference is smaller than the width of the distributions.

\begin{figure*}    
\centering
\includegraphics[width=16cm]{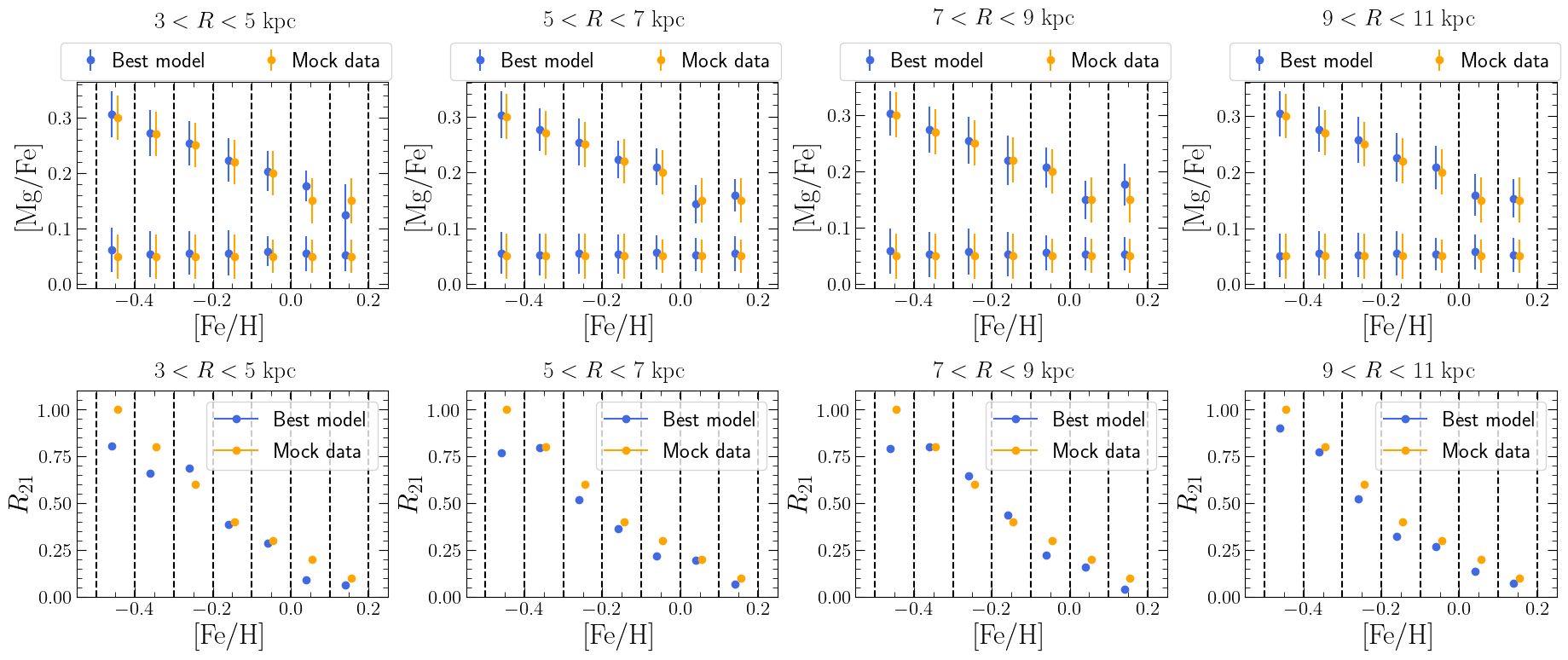}
\caption{Tests of our analysis procedure on a mock catalogue constructed with $|z|$ and age selection
effects like those of APOGEE.  
Upper panels compare model (blue) and data (orange) for the mean value and dispersion of [Mg/Fe] as a function of [Fe/H]. Lower panels show the ratio between the number of stars in the high-[Mg/Fe] and low-[Mg/Fe] sequences, $R_{21}$, integrated over $|z|$ (from $0$ to $\infty$).  }
\label{fig:summary-mock-parameters}
\end{figure*}

\begin{figure*}    
\centering
\includegraphics[width=15cm]{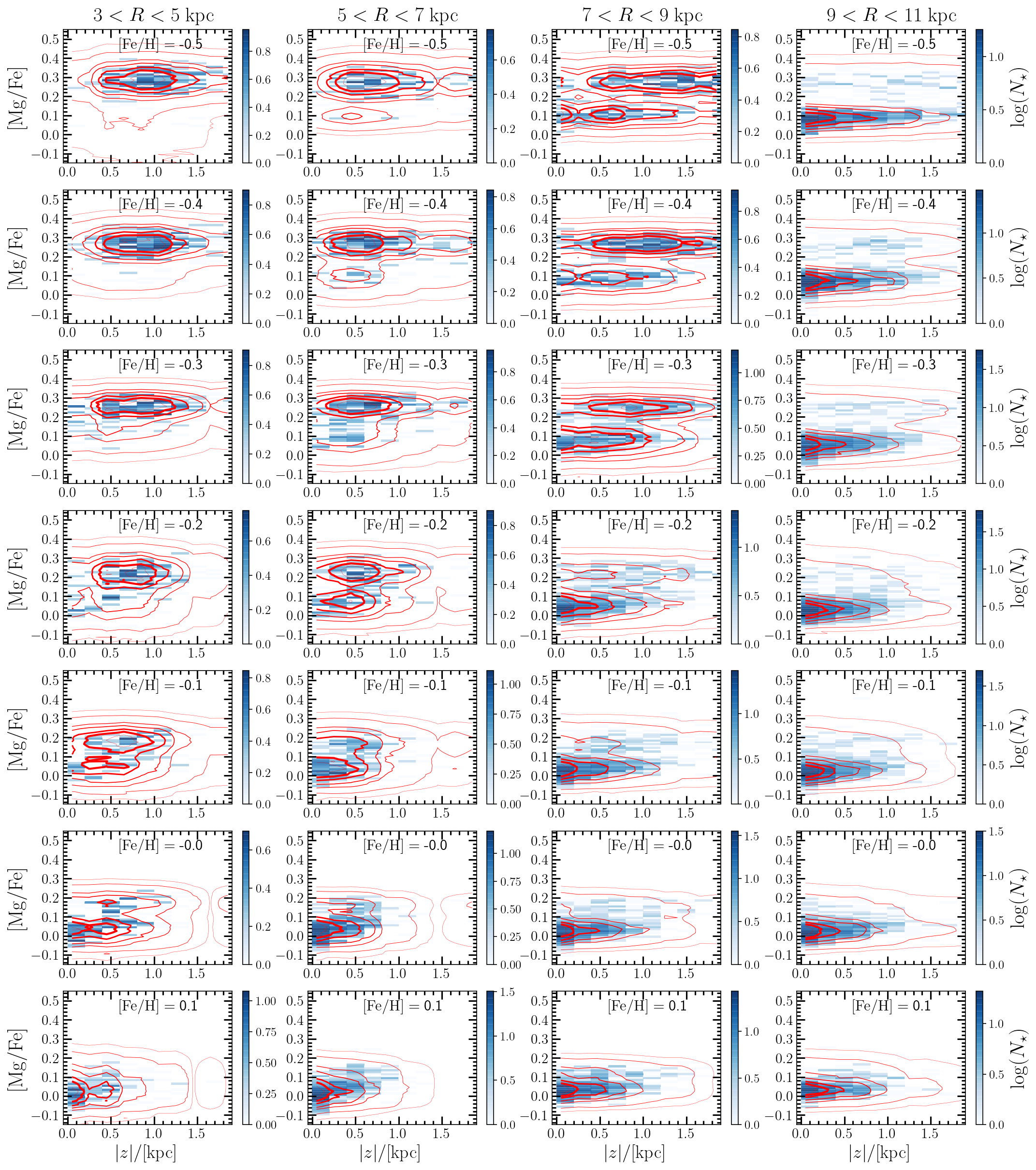}
\caption{Distribution of stars in the $\mgfe-|z|$ plane in ranges of [Fe/H] (rows) and Galactocentric
radius (columns), as labeled.  Our sample of APOGEE-DR16 stars is shown by the colour-coded two-dimensional histograms.  
Red contours show our best fitting model with selection effects included, representing densities normalized to the maximum value in the $(\feh,R)$ bin of
$0.1$, $1$, $10$, $20$, $40$, $60$, and $80$ per cent. }
\label{fig:2d-histograms}
\end{figure*}

\begin{figure}    
\centering
\includegraphics[width=4.5cm]{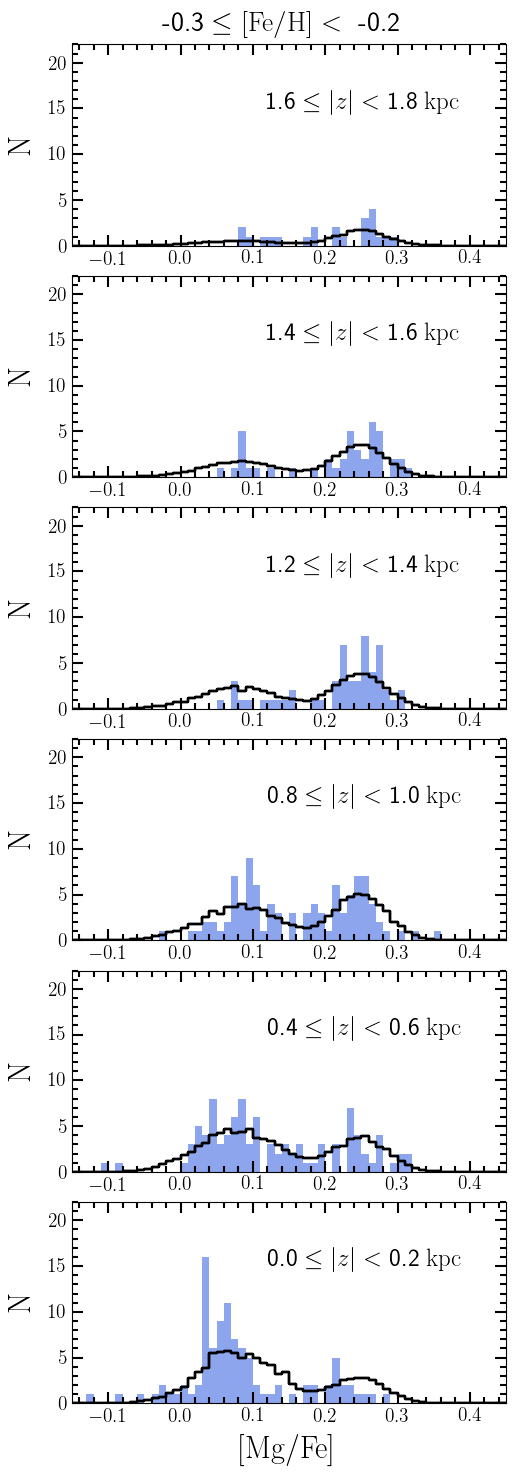}
\caption{The observed APOGEE-DR16 distribution of [Mg/Fe] at the Solar circle (blue histograms) compared to our best-fitting model (black curves), for stars with iron abundances $-0.3 \le \text{[Fe/H]} < -0.2$. Different rows show different $|z|$ bins as labeled, with alternate bins omitted for brevity.  }
\label{fig:distributions-7-9}
\end{figure}

\begin{figure*}    
\centering
\includegraphics[width=17.5cm]{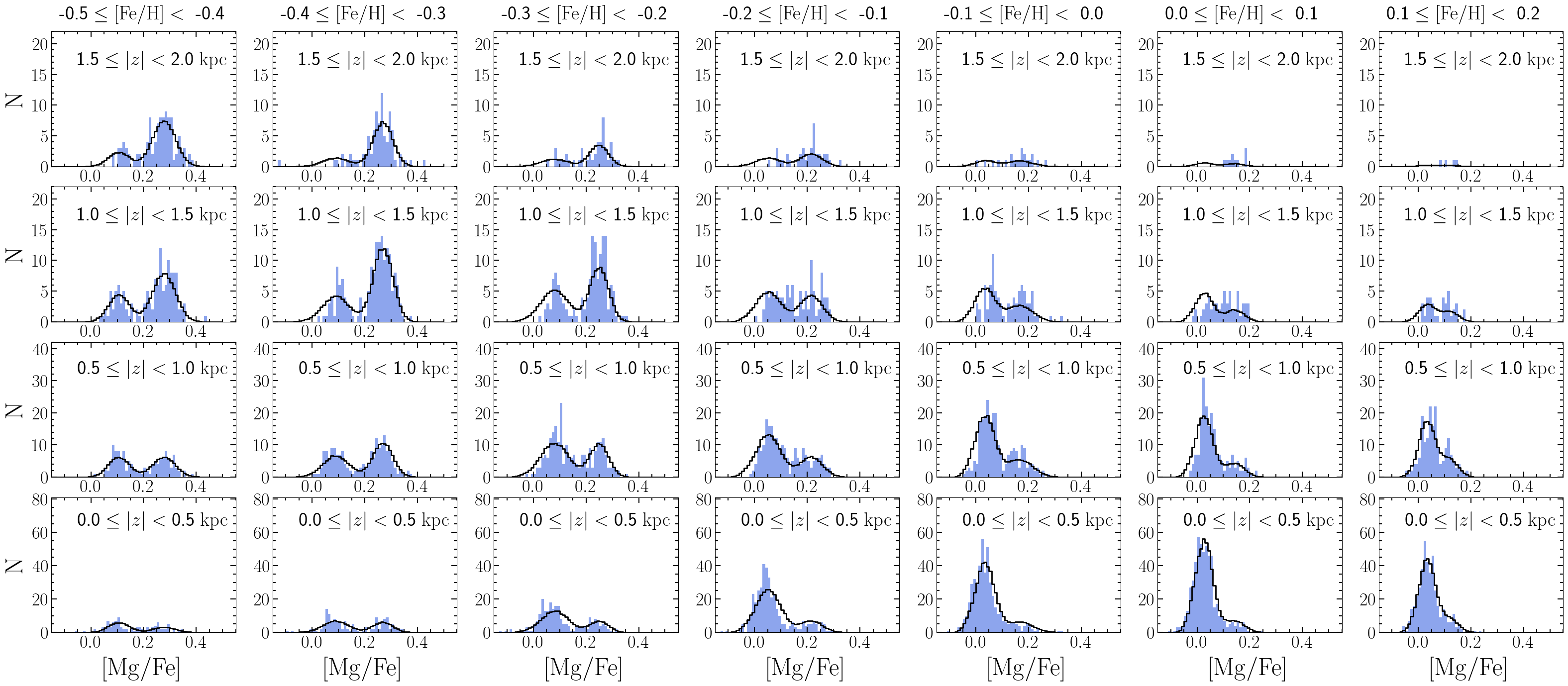}
\caption{Similar to Fig. \ref{fig:distributions-7-9}, but showing all bins of [Fe/H] (columns) and larger ranges of $|z|$ (rows), all at $R=7-9\;\text{kpc}$.  }
\label{fig:comparison_iron}
\end{figure*}

\begin{figure*}    
\centering
\includegraphics[width=10.5cm]{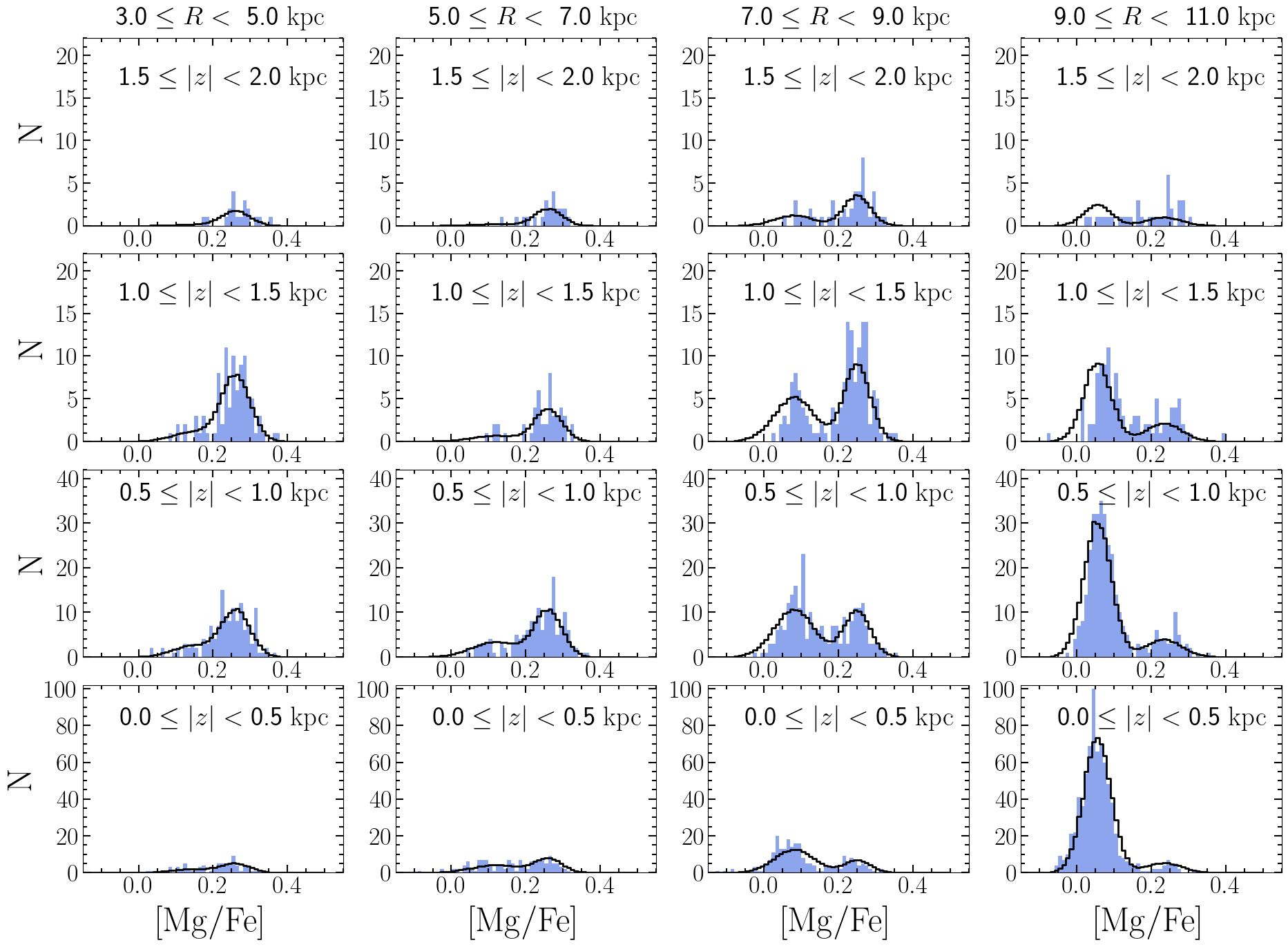}
\caption{Similar to Fig. \ref{fig:comparison_iron}, but showing a range of radial bins (columns), all for the metallicity bin $-0.3 \le \text{[Fe/H]} < -0.2$.  }
\label{fig:comparison_radius}
\end{figure*}

\begin{figure}    
\centering
\includegraphics[width=6.5cm]{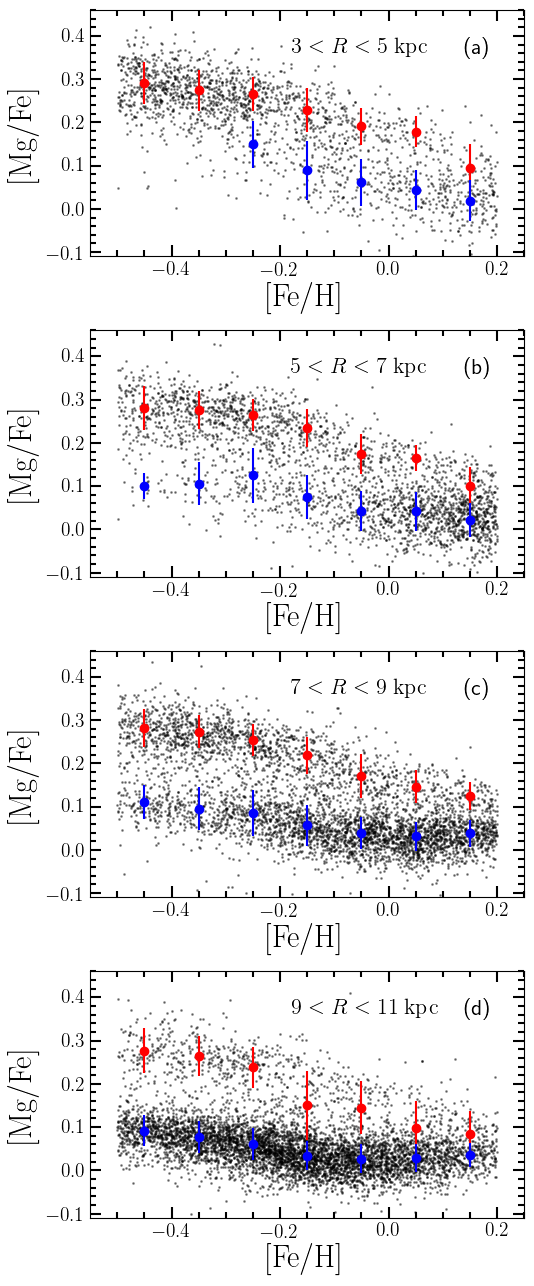}
\caption{The [Mg/Fe]-[Fe/H] relation at different ranges of $R$, as labeled, for all stars with $|z| < 2\;\text{kpc}$ in our APOGEE sample (black points). The red filled circles with error bars show the results of our model for the high-[Mg/Fe] component, whereas the blue filled circles correspond to the low-[Mg/Fe] component. The error bars correspond to the estimated scatter including abundance measurement error (which is small compared to the intrinsic scatter).  }
\label{fig:classical-figure}
\end{figure}

\begin{figure}    
\centering
\includegraphics[width=6.5cm]{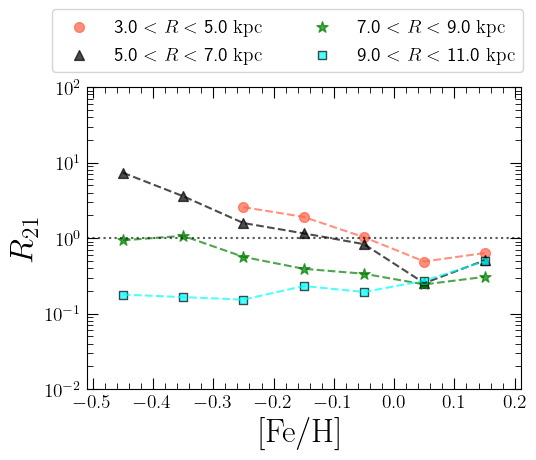}
\caption{The ratio, $R_{21} = n_{2} z_{2} / ( n_{1} z_{1}  )$ between the integrated number of stars in the high-[Mg/Fe] and low-[Mg/Fe] sequences, as predicted by our model as a function of [Fe/H]. Different colors and symbols correspond to different ranges of Galactocentric distance. }
\label{fig:ratio}
\end{figure}

\begin{figure*}    
\centering
\includegraphics[width=15cm]{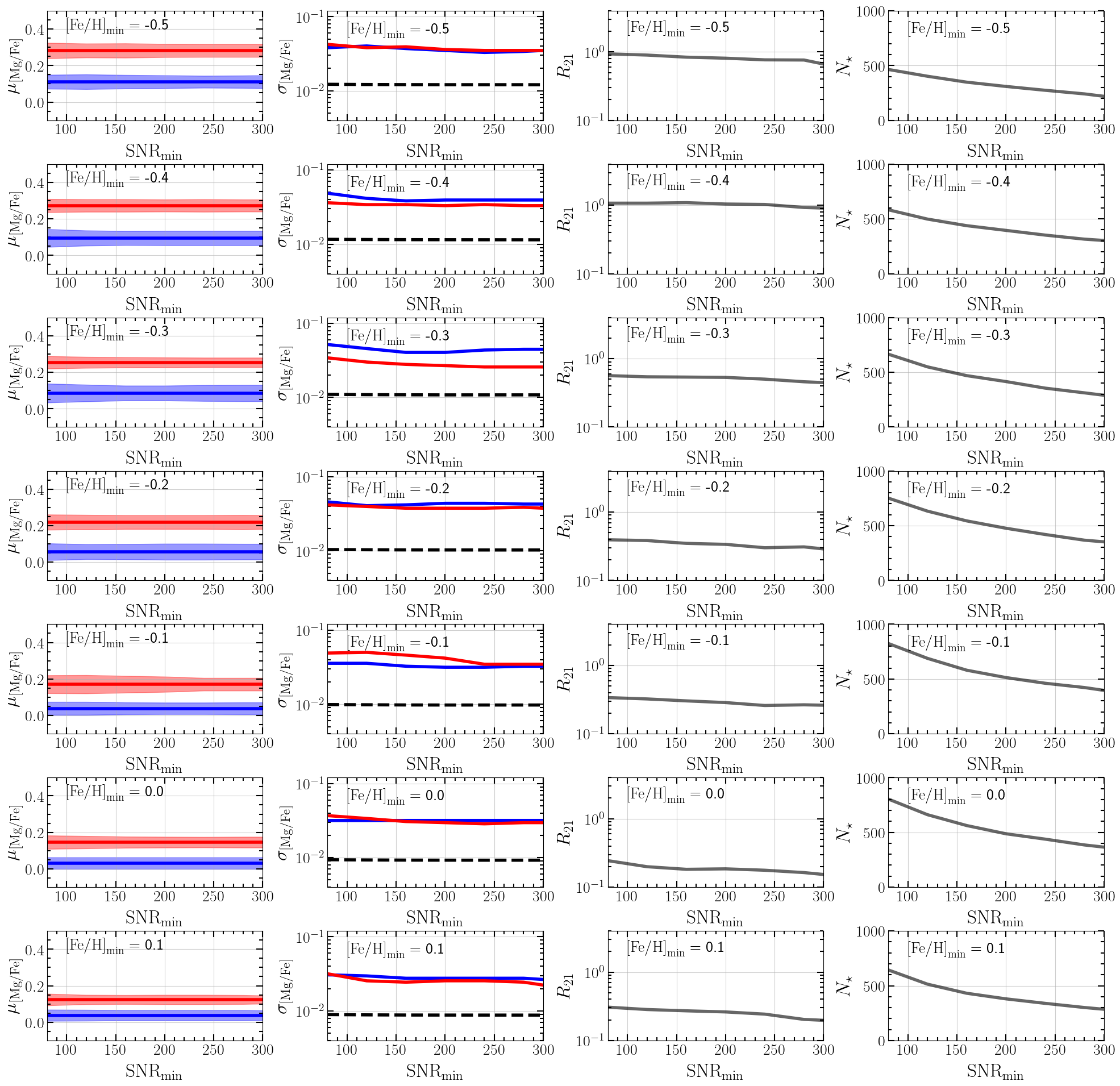}
\caption{Dependence of inferred model parameters on the threshold SNR of the observational sample, for $R=7-9\kpc$.  In each [Fe/H] bin (rows, as labeled) we hold the means $\mu_1$ of the low-$\alpha$ sequence (blue) and $\mu_2$ of the high-$\alpha$ sequence (red) fixed to the values found for SNR$_{\rm min}=80$, as shown in the left column. The second column shows the intrinsic dispersions (solid curves) computed by subtracting the median observational errors (dashed curves) from the best-fit total dispersions.  The third column shows the fitted ratio $R_{21}$ of high-$\alpha$ stars to low-$\alpha$ stars.  The fourth column shows the number of sample stars above the threshold SNR
with $|z| \leq 2\kpc$. 
}
\label{fig:varSNR}
\end{figure*}

\begin{figure*}    
\centering
\includegraphics[width=15cm]{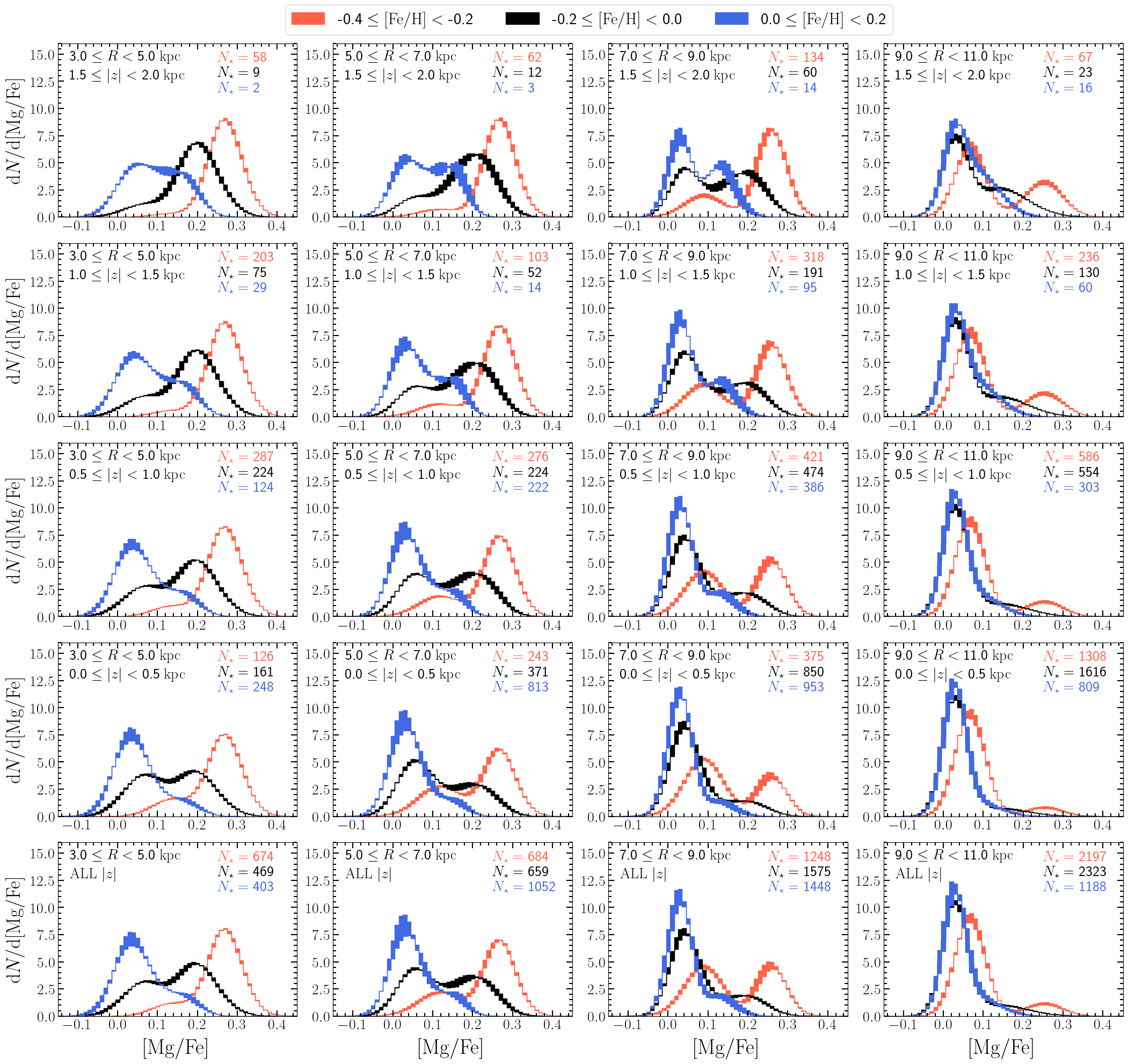} 
\caption{The inferred intrinsic distributions of [Mg/Fe] in three ranges of [Fe/H], in ranges of Galactocentric radius $R=3$-5, 5-7, 7-9, and 9-11 kpc (left to right).  Histograms in the bottom row are integrated over all $|z|$, while rows above show 0.5-kpc ranges of $|z|$ as labeled.  Red, orange, and green curves show metallicity ranges centred on [Fe/H]$=-0.3$, $-0.1$, and $+0.1$, respectively. The finite vertical width of the histograms shows the impact of changing from our standard sample temperature cutoff
 $T_{\text{eff,max}}=4600\;\text{K}$, which excludes red clump stars, to $T_{\text{eff,max}}=4800\;\text{K}$, which includes some red clump stars. In each panel, $N_{\star}$ represents the total number of stars as observed by APOGEE-DR16 in the $(R,|z|,\feh)$ ranges corresponding the histogram of the same colour, by assuming $T_{\text{eff,max}}=4600\;\text{K}$. }
\label{fig:intrinsic-hist}
\end{figure*}

\begin{figure}    
\centering
\includegraphics[width=4.5cm]{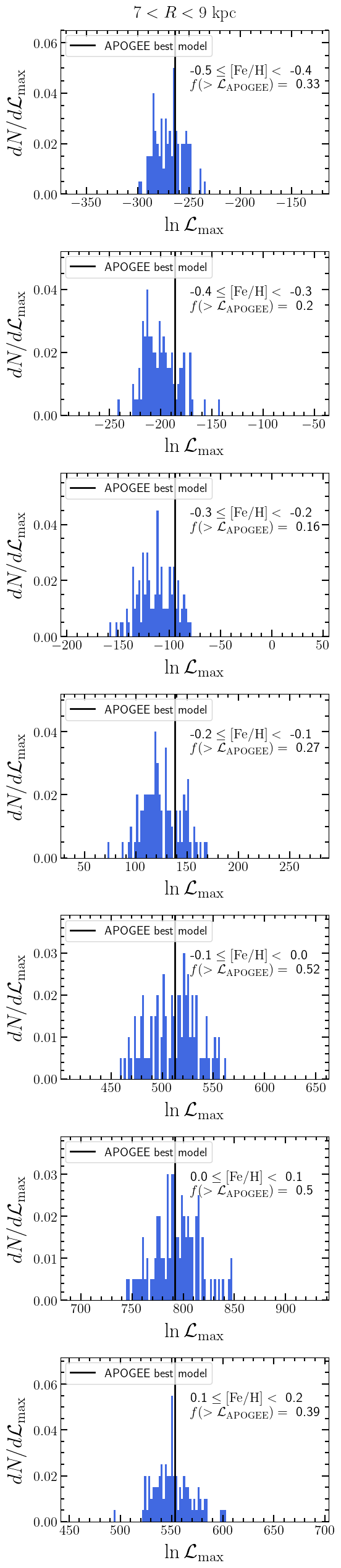}
\caption{The distribution of the log-likelihoods, $\ln\mathcal{L}_{\text{max}}$, of the best-fitting models to our MW mock catalogues for the Solar circle (blue histograms) as a function of [Fe/H] (rows). In each panel, the vertical black line shows the log-likelihood of our best-fitting model to APOGEE-DR16 at the Solar circle, in the given range of [Fe/H]. The model fits the observed APOGEE data as well as it fits mock data sets constructed with the model assumptions, indicating that disagreements are compatible with Poisson fluctuations in the observed star counts. We report the fraction of mock data, $f(>\mathcal{L}_{\text{APOGEE}})$, that yield $\mathcal{L}_{\text{max}}$ higher than our best-fitting model to APOGEE-DR16 at the Solar circle. We set the constant of equation~(\ref{eq:log-likelihood}) to zero when evaluating $\ln\mathcal{L}_{\text{max}}$. }
\label{fig:maxlogL-distribution}
\end{figure}

\begin{figure*}    
\centering
\includegraphics[width=17cm]{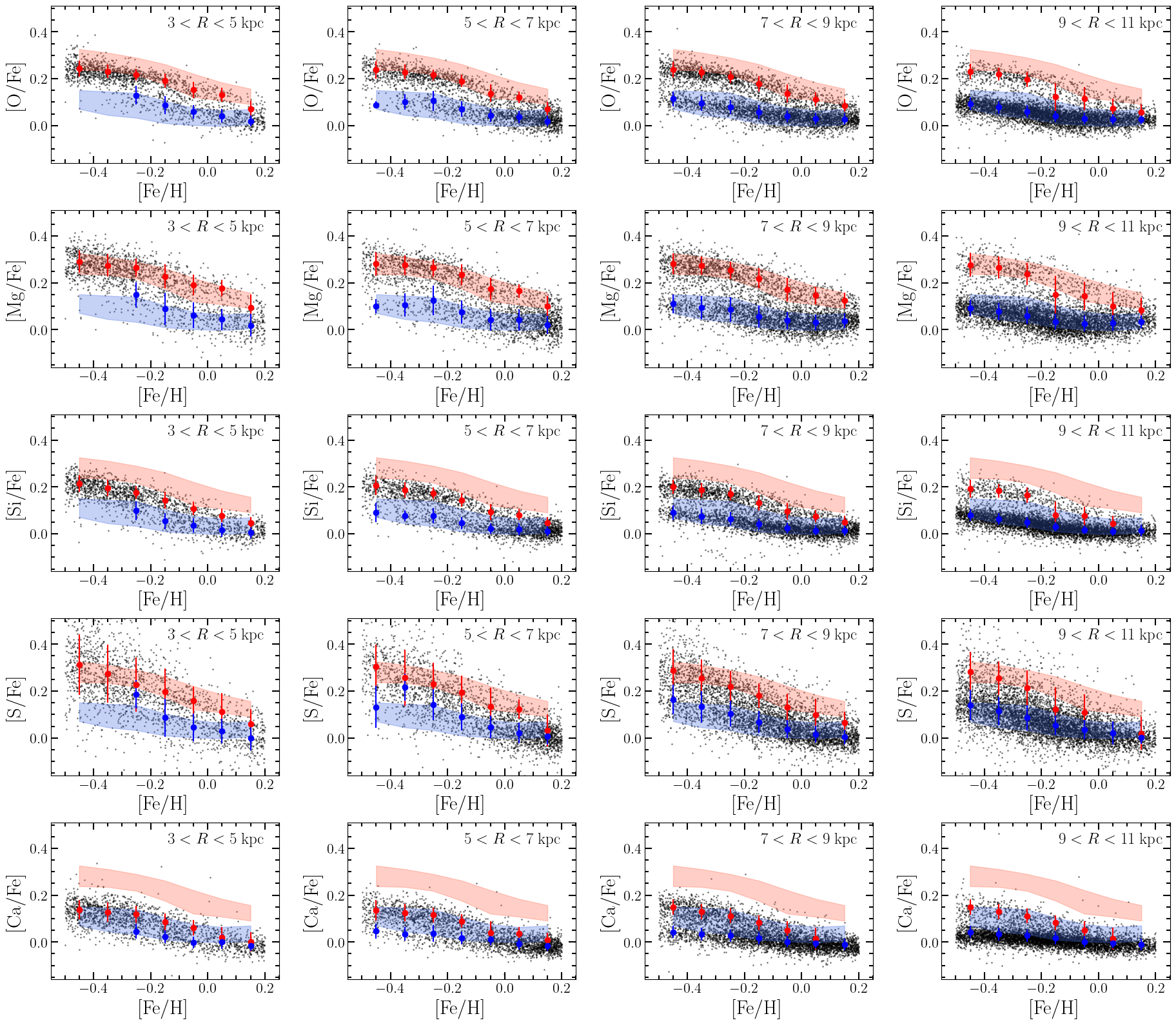}
\caption{The [$\alpha$/Fe]-[Fe/H] relation at different ranges of Galactocentric distance from $R=3$ to $11$ kpc with $\Delta R = 2\,\text{kpc}$ (columns) for five different $\alpha$-elements (rows). From top to bottom, the various rows show a comparison between APOGEE-DR16 observations (black points) and our best-fitting model (filled circles with error bars) for [O/Fe],  [Mg/Fe],  [Si/Fe],  [S/Fe], and  [Ca/Fe]. The red filled circles with error bars correspond to the best-fitting parameters for the high-$\alpha$ sequence, whereas the blue filled circles with error bars correspond to the low-$\alpha$ sequence. The shaded areas are the same in all panels and show the high-$\alpha$ (in red) and low-$\alpha$ (in blue) sequences as determined by our best-fitting model for [Mg/Fe]-[Fe/H] at the Solar circle ($7 \le R < 9 \, \text{kpc}$). }
\label{fig:alpha-feh-2}
\end{figure*}

\section{Model and methods} \label{section:model}

\subsection{The Model} \label{sec:model}

We analyze each bin of [Fe/H] and $R$ independently. 
We wish to derive the underlying distribution of $p(\afe)$, where $\alpha$ represents an individual $\alpha$-element.
We model $p(\afe)$ as the sum of probability density functions (PDFs) of two populations, each characterised by a mean, $\mu$, and dispersion, $\sigma$. 
These PDFs are denoted as $F_{1}(y | \mu_{1}, \sigma_{1})$ and $F_{2}(y | \mu_{2}, \sigma_{2})$, where $y = \afe$ and $\int{ dy\,F_{1}(y)} = \int{dy\,F_{2}(y)} = 1$, with subscript $1$ and $2$ referring to the low-$\alpha$ and high-$\alpha$ populations, respectively.
In this work we assume that both PDFs are Gaussians, namely
\begin{equation}
\label{eq:gaussian}
    F(y | \mu, \sigma) = \left(2\pi\sigma^2\right)^{-1/2}\,e^{ \frac{-(y-\mu)^{2}}{2\sigma^{2}}}.
\end{equation}
We assume that each population has an exponential $z$ distribution (for brevity we write $z$ instead of $|z|$) with scale heights $z_{1}=0.45\;\text{kpc}$ and $z_{2}=0.95\;\text{kpc}$ based on the sub-solar metallicity values suggested by \citet{bovy2016}. The populations thus have
\begin{equation}
\label{eq:z-dependence}
    n(z) = n_{1}\,e^{-z/z_{1}} \;\;\text{and} \;\; n_{2}\,e^{-z/z_{2}}
\end{equation}
\noindent stars per unit $z$. 

In our analysis, we seek to constrain the quantities $n_{1}$, $\mu_{1}$, $\sigma_{1}$, and $n_{2}$, $\mu_{2}$, $\sigma_{2}$, using APOGEE data in narrow bins of $z$. If we take narrow bins $\Delta y \ll \sigma_{1}, \, \sigma_{2}$ and $\Delta z \ll z_{1}, \, z_{2}$, then the number of stars in the bin $y_{i} \leq y < y_{i} + \Delta y$ and $z_{j} \leq z < z_{j} + \Delta z$ is given by 
\begin{equation}
\begin{aligned}
\label{eq:best-model}
\lambda_{ij,\text{model}} = 
\Big[
n_{1}\,F_{1}\big(y_{i} + \frac{\Delta y}{2} \,\Big|\, \mu_{1}, \sigma_{1} \big)\,e^{-(z_{j} + \frac{\Delta z}{2})/z_{1}} + \\ n_{2}\,F_{2}\big(y_{i} + \frac{\Delta y}{2} \,\Big|\, \mu_{2}, \sigma_{2} \big)\,e^{- (z_{j} + \frac{\Delta z}{2})/z_{2}}
\Big] \Delta y \Delta z. 
\end{aligned}
\end{equation} 
Including the correction factors, $f$, for the selection in $T_{\text{eff}}$ and $\log(g)$, the predicted number of stars in the bin $ij$ is 
\begin{equation}
\label{eq:best-model-with-correction}
\begin{aligned}
\lambda_{ij} = Q_{j}\,f(\tau_{i})\Big[n_{1}\,F_{1}\big(y_{i} + \frac{\Delta y}{2} \,\Big|\, \mu_{1}, \sigma_{1} \big)\,e^{-(z_{j} + \frac{\Delta z}{2})/z_{1}} + \\ n_{2}\,F_{2}\big(y_{i} + \frac{\Delta y}{2} \,\Big|\, \mu_{2}, \sigma_{2} \big)\,e^{-(z_{j} + \frac{\Delta z}{2})/z_{2}} \Big] \Delta y \Delta z
\end{aligned}
\end{equation}
where the value of the normalising constant, $Q_{j}$, is chosen so that 
\begin{equation} \label{eq:Kj}
    \sum_{i} \lambda_{ij} = K_{j}, 
\end{equation}
where $K_{j}$ is the number of observed stars in the $(\feh,R)$ bin with
$z_{j}\leq z < z_{j} + \Delta z$.

The values of $f(\tau_{i})$ in equation (\ref{eq:best-model-with-correction}) -- with $\tau_{i}$ corresponding to the ages of the stars in the bin $y_{i} \leq y < y_{i} + \Delta y$ -- are estimated by computing the average of $f(\tau)$ of $N_{\text{rand}}=10$ random ages drawn from a normal distribution, with the mean value being given by the age-[Mg/Fe] relation of \citet[see Fig. \ref{fig:age-mgfe}]{miglio2020} and the dispersion given by the measured asteroseismic errors on the ages (see equation \ref{eq:age-alpha-miglio}). This procedure is done independently for each [Fe/H] bin. 
Although we include scatter to represent the fact that stars of a given [Mg/Fe] have a range of ages, we have checked that choosing all ages from the mean relation with no dispersion makes minimal difference to our results. 

In summary, the free parameters of our model in each bin of $R$ and $\feh$ are $n_{1}$, $\mu_{1}$, $\sigma_{1}$, and $n_{2}$, $\mu_{2}$, $\sigma_{2}$. However, we cannot constrain the true values of $n_{1}$ and $n_{2}$ separately from the observational data, but only their ratio $n_{1}/n_{2}$. 

\subsection{Maximum Likelihood Fitting}

Let $k_{ij}$ denote the observed number of stars in the bin $y_{i} \leq y < y_{i} + \Delta y$ and $z_{j} \leq z < z_{j} + \Delta z$. We assume that the probability of observing $k_{ij}$ stars in the bin $ij$, given a predicted number $\lambda_{ij}$ of stars from the model (see equation \ref{eq:best-model-with-correction}), is described by the Poisson distribution:
\begin{equation}
    p(k_{ij}|\lambda_{ij}) = \frac{\lambda^{k_{ij}}_{ij}\,e^{ -\lambda_{ij} }}{ k_{ij}! } ~.
\end{equation}
The global probability is 
\begin{equation}
    \mathcal{P}( \{ k_{ij} \} | \{ \lambda_{ij} \} ) = \prod_{ij} p( k_{ij}|\lambda_{ij} ) ~,
\end{equation}
\noindent making the log-likelihood 
\begin{equation} \label{eq:log-likelihood}
    \ln \mathcal{L} = \sum_{ij} \ln p(k_{ij}|\lambda_{ij}) = \sum_{ij} k_{ij} \ln \lambda_{ij} - \lambda_{ij} + \text{const.}
\end{equation}
\noindent The constant in equation (\ref{eq:log-likelihood}) depends on the data but not on the parameters, so it can be ignored in finding the maximum likelihood solution.

In our analysis, we consider bins of iron abundances in the range $-0.5 \leq [\text{Fe/H}] < 0.2$ with steps $\Delta\text{[Fe/H]}=0.1$ and bins of $R$ in the range $3 \le R < 11\;\text{kpc}$ with steps $\Delta R = 2\;\text{kpc}$. For each bin of [Fe/H] and $R$, we run our maximum likelihood fitting algorithm using $\Delta y = 0.01\;\text{dex}$ and $\Delta z = 0.2\;\text{kpc}$. 
We optimise $n_{1}$, $\mu_{1}$, $\sigma_{1}$, $n_{2}$, $\mu_{2}$, $\sigma_{2}$ by iterating 1-d grid searches with other parameters being held fixed. Convergence is usually reached well within $\approx10$ iterations for $[\text{Fe/H}] < 0.1$, but it requires $\approx20$ iterations in the range $0.1 \le [\text{Fe/H}] < 0.2$ outside of the Solar circle. We also tested alternative methods based on Markov chain Monte Carlo sampling (\textsc{emcee} package; \citealt{emcee2013}), obtaining similar results.

\subsection{Mock Catalogues} \label{sec:mock-catalogues}
We validate our analysis procedure by applying it to mock catalogues. 
First we create \textit{intrinsic population lists} for the high-$\alpha$ and low-$\alpha$ populations
as follows: 
\begin{enumerate}
    \item We assume that the two populations of artificial stars have a total number $N_{1}=10^{6}$ for the low-$\alpha$ sequence, and $N_{2}= R_{21} \times N_{1}$ for the high-$\alpha$ sequence, with $R_{21}$ being a free parameter. 
    \item We assign a [Mg/Fe] ratio to the artificial stars in the two populations, by randomly sampling two Gaussian functions with different mean values, $\mu_{1}$ and $\mu_{2}$, and dispersions, $\sigma_{1}$ and $\sigma_{2}$. 
    \item We assign a value of $|z|$ to the artificial stars by randomly sampling an exponential function of scale height $z_{1}=0.45\;\text{kpc}$ or $z_{2}=0.95\;\text{kpc}$ for the corresponding population.  The sampling is performed for $0 \le |z| < 10\;\text{kpc}$.
    \item The ages of the artificial stars are assigned from the average age-[Mg/Fe] of \citet[see Fig. \ref{fig:age-mgfe}]{miglio2020}, by using the [Mg/Fe] ratios of the artificial stars from point (ii) above. For each age, $\tau$, we tabulate $f(\tau)$, as computed with equation (\ref{eq:corr}). 
    \end{enumerate}

We choose parameters for our mock catalogue such that it qualitatively resembles our eventual findings
for the APOGEE data.  For the low-$\alpha$ population we take
\begin{equation*}
\mu_{1} = 0.05
\end{equation*}
at all [Fe/H] and
\begin{equation*}
    \begin{aligned}
    \sigma_{1} = [ 0.04,\, 0.04,\, 0.04,\, 0.04,\, 0.03,\, 0.03,\, 0.03 ]
    \end{aligned}
\end{equation*} 
in seven $\Delta\feh=0.1$ bins spanning $-0.5$ to $+0.2$.
For the high-$\alpha$ population we take 
\begin{equation*}
    \mu_{2} = [0.3,\, 0.27,\, 0.25,\, 0.22,\, 0.2,\, 0.15,\,0.15 ]
\end{equation*}
in the same [Fe/H] bins and
\begin{equation*}
    \sigma_{2} = 0.04
\end{equation*} 
in all bins.  
We adopt a number ratio between the two populations (integrated over $|z|$) of
\begin{equation*}
    R_{21} = \frac{n_{2} z_{2}}{n_{1} z_{1}} = [1.0,\, 0.8,\, 0.6,\, 0.4,\, 0.3,\, 0.2,\, 0.1]
\end{equation*}
in the seven [Fe/H] bins.

In each bin of $\Delta R = 2\kpc$ and $\Delta\feh=0.1$, we divide the artificial stars of each 
population into narrow $|z|$ bins of width $0.2\kpc$.  For each such bin we draw stars from the
intrinsic lists with probability proportional to $f(\tau)$.  
 This procedure is repeated until the new list has the the same number of stars, $N([\text{Fe/H}],R,|z|)$, as in APOGEE-DR16 with our reference selection assumptions. 
We apply our analysis to the mock catalogue to determine the values of $\mu_{1}$, $\sigma_{1}$, $\mu_{2}$, $\sigma_{2}$, $R_{21}$, testing that our procedure recovers 
the original parameters assumed to create the intrinsic distributions. 

The results of our analysis are shown in Fig. \ref{fig:summary-mock-parameters}, in which the upper panels show our results for the variation of the mean value and dispersion of [Mg/Fe] as a function of [Fe/H] and the lower panels show our results for the ratio between the number of stars in the high-[Mg/Fe] and low-[Mg/Fe] sequences, $R_{21}$, integrated in $|z|$ from $0$ to $\infty$. The largest discrepancies between the best-fitting model and mock data are found for $R_{21}$ at the lowest [Fe/H] bins, as well as for $\mu_{2}$ (mean [Mg/Fe] of the high-$\alpha$ sequence) at the highest [Fe/H] bins, corresponding to $0.1\le \text{[Fe/H]} < 0.2$.  Even these differences are small, however, and we conclude that our procedure can accurately recover the intrinsic distributions of $\afe$ in the presence of $|z|$ and age selection effects like those in APOGEE.

\section{The Intrinsic Distributions of [$\alpha$/F\texorpdfstring{\MakeLowercase e}{e}]} \label{sec:results}

\subsection{Model Fitting}

We apply our model to fit the observed [Mg/Fe] distributions as observed in a selected sample of stars from APOGEE-DR16, spanning Galactocentric distances $3\le R < 11\;\text{kpc}$ and iron abundances $-0.5 \le \text{[Fe/H]} < 0.2$. The parameters of our best fitting model are reported in Table 1. The values of $\sigma$ in the table correspond to the intrinsic dispersion, defined as the quadrature difference between the fitted total scatter and the median [Mg/Fe] measurement error reported by APOGEE for the low-$\alpha$ and high-$\alpha$ stars in each [Fe/H] bin. 

Fig. \ref{fig:2d-histograms} compares the 2-d distribution of stars in [Mg/Fe] vs. $|z|$ in each bin of [Fe/H] (rows) and $R$ (columns), with the APOGEE counts shown as 2-d histograms and the best-fit model predictions as contours. The model reproduces the changes in the observed distribution across the full range of $R$ and metallicity. Fig. \ref{fig:distributions-7-9} illustrates our fitting procedure more quantitatively, showing predicted and observed distributions of [Mg/Fe] in narrow bins of $|z|$, for $R=7-9\;\text{kpc}$ and $-0.3\le\text{[Fe/H]}<-0.2$. This bin width ($\Delta z = 0.2$) is used for likelihood fitting via equation (\ref{eq:best-model-with-correction}) and (\ref{eq:log-likelihood}). Fig. \ref{fig:comparison_iron} shows comparisons for all [Fe/H] bins with wider bins of $\Delta z = 0.5\;\text{kpc}$. Here we integrate over $\Delta z$ to obtain the model predictions rather than evaluating equation (\ref{eq:best-model-with-correction}) at the bin center. Fig. \ref{fig:comparison_radius} shows all bins of $R$ for $-0.3\le\text{[Fe/H]}<-0.2$. Agreement for other metallicity bins is comparable. Although the functional form of the model is intrinsically smooth, for the comparisons to APOGEE we apply the age selection effects with random draws from $f(\tau | [\alpha/\text{Fe}])$, so the predicted histograms in these figures are not perfectly smooth.

Figs. \ref{fig:2d-histograms}-\ref{fig:comparison_radius} demonstrate good visual agreement between predicted and observed distributions. In Section \ref{sec:log-L-distrib} below, we show quantitatively that the level of agreement is consistent with expectations if the model correctly describes the true distributions up to Poisson fluctuations.

\subsection{Model Parameters and Bimodality in [Mg/Fe]}

Fig. \ref{fig:classical-figure} superposes our best-fit values of $\mu_{1}\pm\sigma_{1}$ and $\mu_{2}\pm\sigma_{2}$ on the [Mg/Fe]-[Fe/H] abundance distribution of all sample stars with $|z|<2\;\text{kpc}$ in the four Galactocentric radius bins. Except at the highest [Fe/H] values, the means $\mu_{1}$ and $\mu_{2}$ differ by more than $\sigma_{1}+\sigma_{2}$, an indication of two distinct sequences at all $R$. As shown previously by H15 and W19, the high-$\alpha$ and low-$\alpha$ sequences have similar locations at all $R$, though in detail our inferred values of $\mu$ depend on $R$, and a fit enforcing constant values at all $R$ is significantly worse.

Physically, the mean value of [Mg/Fe] for stars on the low-[Fe/H] plateau of the high-$\alpha$ sequence should represent the ratio of the IMF-averaged yield of Mg and Fe from massive stars, dying as core-collapse SNe. At $\text{[Fe/H]}\approx-0.5$, we find $\mu_{2} \approx 0.29$ for $3\leq R < 5\;\text{kpc}$, and $\mu_{2} \approx 0.28$ for $R \ge 5\;\text{kpc}$. As [Fe/H] increases, the value of $\mu_{2}$ decreases because of the increasingly large contribution of Fe from Type Ia SNe. As discussed in the introduction, the origin of the low-[$\alpha$/Fe] sequence is a matter of debate, including the extent to which it represents an evolutionary sequence at all.   

Fig. \ref{fig:ratio} shows the ratio $R_{21}$ of stars in the high-$\alpha$ and low-$\alpha$ populations. Those ratios are integrated over $|z|$ and corrected for the age-dependent selection. They depend on $R$ and [Fe/H], but the $z$-dependence of the observed ratio is a consequence of the differing scale heights $z_{2}=0.95\;\text{kpc}$ and $z_{1}=0.45\;\text{kpc}$. As one would guess from the uncorrected stellar distributions, the high-$\alpha$ population is more important at low $\text{[Fe/H]}$ and small $R$. At the Solar circle, the populations are equal for $-0.5 \le \text{[Fe/H]}<-0.3$, and the low-$\alpha$ population is larger at $\text{[Fe/H]}>-0.3$. For $R=9-11\;\text{kpc}$, the high-$\alpha$ population is subdominant at all [Fe/H] examined here. Varying $z_{2}$ has a little impact on our findings, whereas varying $z_{1}$ has an impact on $R_{21}$ at low [Fe/H], such that if $z_{1}$ is diminished, then $R_{21}$ at low [Fe/H] also diminishes in the fit to compensate for that variation.

Fig. \ref{fig:varSNR} shows how the inferred values of the intrinsic dispersion and the $R_{21}$ population ratio at the Solar circle change when increasing the sample's minimum SNR from $\text{SNR}_{\text{min}}=80$ to $300$. We keep $\mu_{1}$ and $\mu_{2}$ fixed to the values at $\text{SNR}_{\text{min}}=80$ (see Table 1), but they do not change much if we leave them free. 
The number of stars in each [Fe/H] bin drops by a factor of $2-2.5$ as we increase $\text{SNR}_{\text{min}}$ from $80$ to $300$, but the values of $\sigma_{1}$, $\sigma_{2}$, and $R_{21}$ show only small variations. Dotted lines in the second column show the median observational errors in [Mg/Fe] reported by APOGEE as a function of the SNR threshold. The median error is almost independent of $\text{SNR}_{\text{min}}$.  The inferred intrinsic scatter, obtained by subtracting this median error in quadrature from the best-fit total dispersion, is a factor $\sim 3-4$ larger than the observational error and insensitive to the SNR threshold. In principle, scatter in the observed sequences could arise from systematic errors in the abundance determinations that vary from star to star.  Our use of a narrow range of stellar parameters in our sample is designed to limit this effect.  As a further check, we have examined histograms of [Mg/Fe] for the cooler and hotter stars of our sample ($\Teff = 4200-4400\;{\rm K}$ vs.\ $\Teff = 4400-4600\;{\rm K}$) and the lower and higher $\logg$ stars of our sample ($1-1.75$ vs.\ $1.75-2.5$).  We find no clear offsets of [Mg/Fe] between these data subsets, and any shifts of histogram peaks are small compared to the widths of the high-$\alpha$ and low-$\alpha$ distributions.  

From the insensitivity to SNR$_{\rm min}$ and the test on data subsets,
we conclude that our model values provide robust estimates of the intrinsic spread of [Mg/Fe] at fixed [Fe/H] for the high-$\alpha$ and low-$\alpha$ populations. These intrinsic dispersions, typically $0.03-0.05\;\text{dex}$ (see Table 1), are themselves a significant constraint on GCE models. 

The best-fit parameter values in 
Table~1,
and for other $\alpha$-elements in 
Table~2
(see Section \ref{sec:other-alpha}), constitute the principal results of this paper.  Fig.~\ref{fig:intrinsic-hist} plots our findings for the conditional distributions $p(\mgfe \,|\, \feh)$ in ranges of $R$ and $|z|$ for the three metallicity bins centred on $\feh=-0.3$, $-0.1$, and $+0.1$.  These model histograms are corrected for the age-dependent and $|z|$-dependent selection effects in APOGEE, so they show the distributions expected if one could observe a random subset of all long-lived stars (lifetime greater than the age of the Galactic disk) in the indicated $R$ and $|z|$ range.  To quantify the uncertainty due to the dependence of the predicted correction factors on $T_{\text{eff}}$ (see Fig. \ref{fig:correction-factor}), we compare the model distributions obtained for our reference selection with $T_{\text{eff,max}}=4600\;\text{K}$ (which removes red clump stars) to those obtained with  $T_{\text{eff,max}}=4800\;\text{K}$ (which includes some red clump stars), for which we recompute the correction factors.

The bottom row of Fig.~\ref{fig:intrinsic-hist} shows results for the full disk populations integrated over $|z|$, the best representation of the relative importance of the high-$\alpha$ and low-$\alpha$ populations at a given $R$.  Our key finding is that the bimodality of the $\mgfe$ distribution at sub-solar $\feh$ in the solar neighborhood is a genuine, intrinsic fea ture of the stellar populations, not an artefact of over-representing the high-$\alpha$ stars.  In the $|z|$-integrated histogram for $R=7-9\kpc$, the minimum of the $\feh \approx -0.3$ conditional distribution is a factor of three below the surrounding maxima.  The sharpness of bimodality depends on $R$, $|z|$, and $\feh$.  In some ranges, the lower amplitude Gaussian adds a shoulder or asymmetric tail to a distribution with a single maximum.  As expected from previous results (e.g., \citealt{bensby2003,lee2011}; H15), the high-$\alpha$ population is more prominent at larger $|z|$, lower $\feh$, and smaller $R$.  The $R$-dependence is sometimes characterized as a shorter scale length for the ``chemical thick disk'' vs.\ the ``chemical thin disk'' \citep{bovy2016}.  A fully successful model of the Milky Way must reproduce these trends, and the clear bimodality of $\mgfe$ in some ranges, and the intrinsic dispersions of $\mgfe$ in the high-$\alpha$ and low-$\alpha$ populations.

\subsection{Goodness of fit}
\label{sec:log-L-distrib}

From Figs.~\ref{fig:2d-histograms}-\ref{fig:classical-figure} it is clear that our parameterized model provides a good qualitative description of the [Mg/Fe] distributions observed by APOGEE.  To give a quantitative measure of goodness-of-fit, analogous to a frequentist $\chi^2$-test, we create mock data sets that satisfy our model assumptions by construction, then fit them using the same procedure applied to the observational data.  We generate 100 such data sets for each 0.1-dex bin in the range $-0.5\le\text{[Fe/H]}<0.2$.  Each mock catalogue is generated with the procedure outlined in Section \ref{sec:mock-catalogues}, by assuming for $R_{21}$, $\mu_{1}$, $\mu_{2}$, $\sigma_{1}$, and $\sigma_{2}$ the values reported in 
Table 1 
as a function of [Fe/H] for the Solar circle.  We draw numbers of stars in each range of [Fe/H] that match those in our $R=7-9\kpc$ observational sample.  For each mock catalogue we record $\ln\mathcal{L}_{\rm max}$ of the best-fit model (eq.~\ref{eq:log-likelihood} with the constant set to zero).

Fig. \ref{fig:maxlogL-distribution} shows the histogram of $\ln\mathcal{L}_{\rm max}$ values for the 100 mock catalogues in each [Fe/H] bin, as well as the value obtained by fitting the APOGEE data.  In all cases the observational value falls well within the mock catalogue distribution, indicating that the model describes the data to within the level expected from Poisson fluctuations.  In detail, the observed $\ln\mathcal{L}_{\rm max}$ values for all seven [Fe/H] bins fall within the upper 52\% of the mock catalogue distributions, which could be taken as evidence for a slight tension between the model and the data.

\subsection{Other $\alpha$-elements}
\label{sec:other-alpha}

We apply a similar analysis to determine the abundance distributions of the $\alpha$-elements O, Si, S, and Ca, relative to iron, as observed by APOGEE-DR16. In these cases, our analysis makes the following simplifications:
\begin{enumerate}
    \item In the maximum likelihood fitting, the parameter $R_{21}$ as a function of radius and [Fe/H] is held fixed to the best values that we have determined for the [Mg/Fe] distributions (see Fig. \ref{fig:ratio} and Table 1). 
    \item In order to correct for the age-selection effects for element X, the values of $\tau_{i}$ in equation (\ref{eq:best-model-with-correction}) (ages of the stars in the bin $y_{i} \leq y < y_{i} + \Delta y$, where $y_{i}=[\text{X}/\text{Fe}]$) are estimated with the following two-step procedure. \textit{(a)} Firstly, we determine the average $[\text{X}/\text{Fe}]$-[Mg/Fe] relation, by binning the APOGEE-DR16 $[\text{X}/\text{Fe}]$ ratios in this plane and making use of a linear interpolation to determine the average [Mg/Fe] from any given $y_{i}=[\text{X}/\text{Fe}]$. \textit{(b)} Once we determine the average [Mg/Fe] corresponding to $y_{i}$, we use the average [Mg/Fe]-age relation from \citet[see Fig. \ref{fig:age-mgfe}]{miglio2020} to estimate $\tau_{i}$. 
\end{enumerate}
The results of our analysis are shown in Fig. \ref{fig:alpha-feh-2}, in which the different columns for a given row show the [$\alpha$/Fe]-[Fe/H] pattern at different radii. From top to bottom, the various rows show $\alpha$-elements in increasing order of atomic mass, namely [O/Fe], [Mg/Fe], [Si/Fe], [S/Fe], and [Ca/Fe]. The filled circles with the error bars in Fig. \ref{fig:alpha-feh-2} correspond to the mean values and dispersion of the high-$\alpha$ (in blue) and low-$\alpha$ (in red) sequences, whereas the coloured shaded areas highlight the [Mg/Fe]-[Fe/H] relation at the Solar circle ($7\le R < 9\,\text{kpc}$) and are the same in all panels.
In Table 2, we report the values of $\mu_{1}$, $\mu_{2}$, $\sigma_{1}$, and $\sigma_{2}$ as a function of [Fe/H] and $R$ for the various $\alpha$-elements shown in Fig. \ref{fig:alpha-feh-2}. 

The low-$\alpha$ sequence appears similar for all of these elements with the exception of Ca, which is displaced to lower [Ca/Fe].  The separation of the high-$\alpha$ and low-$\alpha$ sequences is smaller for [Si/Fe] and [Ca/Fe] than for [Mg/Fe]; a similar but smaller offset of the high-$\alpha$ sequence is seen for [O/Fe] and [S/Fe].  As discussed by W19, a smaller separation of these sequences would arise if SNIa or another delayed nucleosynthetic source makes a larger contribution to these elements than to Mg.  At the Solar circle, the scatter on the low-$\alpha$ sequence is smaller for O, Si, and Ca than for Mg, while scatter on the high-$\alpha$ sequence is comparable or slightly smaller.  For a given element, the location of the high-$\alpha$ and low-$\alpha$ [X/Fe]-[Fe/H] sequences is only weakly dependent on $R$; the changes with $R$ for a given element are much smaller than the apparent differences from one element to another.  There is a general tendency for the ``knee'' in the high-$\alpha$ sequence to move to lower $\feh$ at large $R$, as predicted in GCE models where the star formation efficiency decreases as a function of $R$ (e.g., \citealt{matteucci1989,boissier1999,cescutti2007,belfiore2019,palla2020}). APOGEE-DR16 incorporates zero-point calibration offsets such that stars in the solar neighborhood with $\feh \approx 0$ have a mean $[\text{X}/\text{Fe}] = 0$ \citep{ahumada2020,jonsson2020}.  However, APOGEE abundances exhibit systematic trends with log$\,g$ at the $\sim 0.05$-dex level \citep{jonsson2020,griffith2020}.  Element-to-element offsets at this level should therefore be treated with caution.

\section{Conclusions} \label{sec:conclusions}

We have used abundance measurements from APOGEE-DR16 to infer the
conditional distributions $p(\xfe\,\big|\,\feh)$ for the $\alpha$-elements
Mg, O, Si, S, and Ca in the range $-0.5 \leq \feh < +0.2$.
Our sample consists of Milky Way disk stars in bins of Galactocentric radius
$R=3$-5, 5-7, 7-9, and 9-11 kpc.  By fitting a model and accounting for
age-dependent and $|z|$-dependent selection effects, we translate the
observed distributions into the intrinsic distributions that would be
found by observing a random sample of all long-lived disk stars.
Our model assumes a double Gaussian $p(\afe)$ in each 0.1-dex $\feh$ bin,
one describing the low-$\alpha$ population with scale height $z_1=0.45\kpc$ and one describing the high-$\alpha$ population with scale-height $z_2=0.95\kpc$.
The means, dispersions, and relative amplitude of the two Gaussians
are determined by maximum likelihood fitting.  Tests on mock catalogues
demonstrate that the method yields reliable parameter determinations
and that the agreement with the APOGEE data is at the level expected
if the model assumptions are correct.  This agreement is illustrated
in Figs.~\ref{fig:2d-histograms}-\ref{fig:classical-figure} and Fig.~\ref{fig:maxlogL-distribution}.

Our principal results consist of the model parameter determinations reported
in 
Tables~1 and~2
and illustrated in
Figs.~\ref{fig:classical-figure}, \ref{fig:ratio}, \ref{fig:varSNR}, and \ref{fig:alpha-feh-2}.
The means, dispersions, and relative amplitudes of these sequences are
quantitative measures that a complete model of Milky Way disk chemical
evolution should reproduce.

The intrinsic $\mgfe$ distributions that we infer are illustrated
in Fig.~\ref{fig:intrinsic-hist}.  In qualitative terms, our most important finding
is that the bimodality of $\afe$ ratios observed near the solar radius
is an intrinsic property of the disk stellar populations, not an
artefact of over-representing thick-disk stars.  In agreement with
previous results (H15 and references therein)
we find that the high-$\alpha$ population is more prominent at
small $R$, low $\feh$, and large $|z|$.  We also find that the separation
of the high-$\alpha$ and low-$\alpha$ sequences is smaller for Si and Ca
than for Mg, O, and S, implying a larger SNIa contribution to these
elements (W19).

\begin{figure*}    
\centering
\includegraphics[width=12cm]{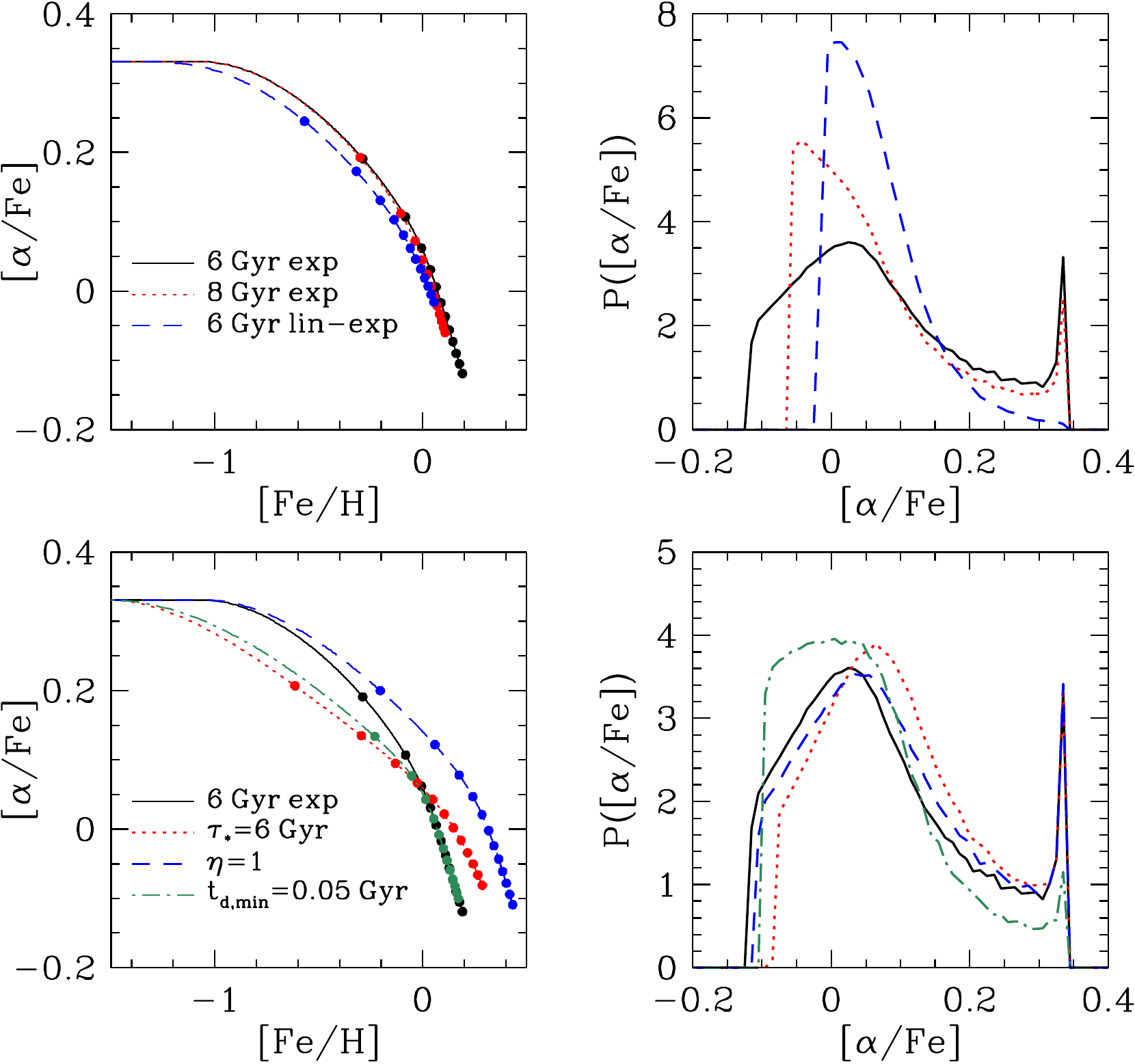} 
\caption{Evolutionary tracks (left) and $\afe$ distributions (right) of one-zone chemical evolution models, computed using the analytic formalism of WAF.  Points in the left panels show $\afe$ and $\feh$ at 1 Gyr intervals (up to 12 Gyr).  Black solid curves show a reference model with outflow efficiency $\eta=2$, star formation efficiency $\tau_*^{-1} = (2\Gyr)^{-1}$, SNIa minimum delay time $t_{d,{\rm min}}=0.15\Gyr$, and an exponentially decaying star formation history with a timescale of 6 Gyr.  In the upper panels, red dotted and blue dashed curves show models with an 8 Gyr exponential or 6 Gyr linear-exponential star formation history, respectively.  In the lower panels, red dotted, blue dashed, and green dot-dashed curves show the effect of changing the star formation efficiency to $\tau_*^{-1} = (6\Gyr)^{-1}$, the outflow efficiency to $\eta=1$, or the minimum delay time to $t_{d,{\rm min}}=0.05\Gyr$.
}
\label{fig:onezone}
\end{figure*}

As context for interpreting these results, we show $\afe-\feh$ tracks and
$\afe$ distribution functions of one-zone chemical evolution models
in Fig.~\ref{fig:onezone}, computed using the analytic methods of WAF.
An individual one-zone model with constant parameters cannot explain
the observed $\afe-\feh$ distribution of disk stars because it predicts
that $\feh$ and $\afe$ evolve monotonically; the $\afe$ distributions
in Fig.~\ref{fig:onezone} are computed over all $\feh$ values, not
narrow bins like those of in our APOGEE analysis.  However, a mixture
of populations evolved along distinct evolutionary tracks at different
Galactocentric radii will exhibit a distribution of $\afe$ at fixed $\feh$
(e.g., \citealt{schoenrich2009,minchev2013,nidever2014}).
A one-zone model in which outflow efficiency grows at late times can
produce a double-valued $\afe-\feh$ track in which the low-$\alpha$
sequence evolves from high $\feh$ to low $\feh$ (WAF), but we do
not explore this possibility here.

Black curves in each panel show results for a fiducial model with
star formation efficiency timescale $\tau_*=2\Gyr$, outflow mass
loading $\eta=2$, and an exponentially declining SFH with $e$-folding
timescale $\tau_{\rm sfh}=6\Gyr$ (see WAF for modeling details).
The SNIa delay time distribution is a sum of exponentials designed
to mimic a $t^{-1.1}$ power law with a minimum delay time
$t_{d,{\rm min}}=0.15\Gyr$.  The $\afe$ distribution has a narrow
peak at the plateau value set by the CCSN yield ratio, then 
a broad peak centred near $\afe \approx 0$.  Increasing the star-formation
timescale to 8 Gyr (upper panels, red curves) 
has little effect on the high-$\alpha$
peak, but the low-$\alpha$ peak becomes narrower.  For a constant SFR
(not shown) the model evolves quickly to equilibrium abundances, and
the $\afe$ distribution is sharply peaked at the Solar ratio.
Blue curves in the upper panels show a linear-exponential 
${\rm SFR} \propto t e^{-t/\tau_{\rm sfh}}$, which evolves more slowly
than the corresponding exponential model.  With this SFH, the
high-$\alpha$ peak disappears entirely because fewer stars are
formed at early times.

The lower panels of Fig.~\ref{fig:onezone} show the impact of changing
other model parameters with the SFH (6 Gyr exponential) held fixed.
Decreasing the star formation efficiency, to $\tau_*^{-1} = (6\Gyr)^{-1}$,
moves the knee of the $\afe-\feh$ track to lower metallicity but only
slightly alters the $\afe$ distribution.  Similarly, a model with
reduced outflow efficiency, $\eta=1$, evolves to higher $\feh$ but
has nearly the same $\afe$ distribution.  Reducing 
$t_{d,{\rm min}}$ to 0.05 Gyr suppresses the high-$\alpha$
peak (and shifts the knee to lower $\feh$) because of the more
rapid onset of SNIa iron enrichment, and the low-$\alpha$ peak
becomes broad and flat.

While we reserve tests of theoretical models to future work, we note 
that the clear minima in the $\afe$ distributions found in many zones
at sub-solar $\feh$ pose a challenge to models that explain the
$\afe-\feh$ distribution purely in terms of radial mixing of populations
that follow separate evolutionary tracks in this space.
Because one-zone models with smooth star formation histories generically
produce $\afe$ distributions that are strongly peaked at low
values, it is difficult to construct a superposition of such distributions
with two well separated maxima of comparable strength.
The \cite{sharma2020} model may appear to represent a counter-example,
but \cite{sharma2020} do not demonstrate that the chemical evolution
tracks assumed in this model can be produced self-consistently by
an underlying star formation and enrichment history.  Radial mixing
is almost certainly an important ingredient in Galactic chemical evolution,
but we suspect that explaining the bimodality of $\afe$ will also require
some form of ``discontinuous'' evolution, whether that be the 3-phase
star formation history envisioned in the 2-infall scenario
\citep{chiappini1997,spitoni2019,palla2020}, the clumpy bursts of star formation
proposed by \citet{clarke2019}, the sharp change of outflow efficiency
suggested by WAF, the unusual gas accretion history found in a small
fraction of simulated galaxies \citep{grand2018,mackereth2018},
or an early merger event that resets the metallicity of the ISM 
\citep{buck2020,vincenzo2020}.

In agreement with the findings for [O/Fe] by \cite{bertran2016}, we
infer intrinsic scatter of $\mgfe$ at the $\sim 0.04$-dex level on both
the high-$\alpha$ and low-$\alpha$ sequences, and comparable values
for other elements.  The implied fractional scatter of ($\alpha$/Fe) ratios
is thus $\sim 10\%$.  Predicting scatter of abundance ratios requires
a detailed model for the mixing of supernova products in ISM gas and
subsequent star formation (e.g., \citealt{krumholzting2018}),
and we do not know of quantitative predictions of $\afe$ scatter
in the metallicity range examined here.  Stochastic sampling of the
stellar IMF is probably not sufficient: in round numbers, enriching
a gas mass $M_g$ to $0.5Z_\odot$ requires 
$N_{\rm SN} \sim 10^4 (M_g/5\times 10^6 M_\odot)$ core-collapse
supernovae, so order unity fluctuations in (X/Fe) from one supernova
to another would average down to $\sim 0.01(M_g/5\times 10^6)^{-1/2}$ rms
fluctuations in the enriched stellar population.  
On the low-$\alpha$ sequence, the scatter could
plausibly arise from a several-Gyr range of stellar ages that leads to 
star-by-star variations in the ratio of SNIa-to-CCSN enrichment
(see Fig.~\ref{fig:onezone}).  This explanation on its own may be
insufficient for the high-$\alpha$ sequence because populations
evolve away from the plateau so quickly.  However, superposition
of populations that have different knees in their $\afe-\feh$ 
evolution will produce scatter at moderately sub-solar $\feh$ 
from the mix of stars that lie varying distances below the plateau
(see Fig.~\ref{fig:onezone}, lower left).
Larger scatter could arise from superposing populations that have
different $\afe$ plateau values, which would require systematic
differences in the supernovae associated with these populations.
Empirical clues to the origin of scatter can be derived by examining
element-by-element correlations of star-by-star deviations from 
mean trends, and correlations of these deviations with stellar age
and kinematics.  We will undertake such studies in future work.

Spectroscopic surveys like APOGEE, GALAH, RAVE, and the Large Sky Area Multi-Object Fibre Spectroscopic Telescope (LAMOST; e.g. \citealt{Xiang2019,wheeler2020}) have extended the study of element ratio distributions from the solar neighborhood or other selected regions
to comprehensive maps across much of the Galaxy, with the precision
and well characterized selection needed to resolve such questions as
bimodality and intrinsic scatter (H15; W19; \citealt{griffith2020,nandakumar2020}.  
These empirical distributions 
encode information about many aspects of Galactic evolution and 
enrichment physics, including accretion and star formation history,
mergers and perturbations, gas mixing and gas flows, radial migration
of stars, and supernova nucleosynthesis.  They present powerful tests
for increasingly sophisticated models of galactic chemical evolution and
for hydrodynamic simulations of galaxy formation.


\begin{table*}
\centering \label{table1}
\begin{tabular}{c|c|c|c|c|c|c|c}
\hline
\hline
\multicolumn{8}{c}{$\mathbf{3\le R < 5 \;\text{\bf kpc}}$} \\
  $\text{[Fe/H]}_{\text{min}}$ & $-0.5$ & 
  $-0.4$ & 
  $-0.3$ & 
  $-0.2$ & 
  $-0.1$ & 
  $0.0$ & 
  $0.1$   \\[3pt]
\hline

$\mu_{1}$ & $-$ & $-$ & $0.149$ & $0.089$ & $0.061$ & $0.044$ & $0.019$ \\
$\sigma_{1}$ & $-$ & $-$ & $0.052$ & $0.067$ & $0.053$ & $0.044$ & $0.045$  \\
$\mu_{2}$ & $0.291$ & $0.274$ & $0.265$ & $0.228$ & $0.191$ & $0.178$ & $0.095$  \\
$\sigma_{2}$ & $0.046$ & $0.045$ & $0.036$ & $0.049$ & $0.041$ & $0.034$ & $0.055$  \\
$R_{21}$ & $18.357$ & $17.911$ & $2.574$ & $1.912$ & $1.030$ & $0.491$ & $0.635$  \\[3pt]

\hline
\hline

\multicolumn{8}{c}{$\mathbf{5\le R < 7 \;\text{\bf kpc}}$} \\
  $\text{[Fe/H]}_{\text{min}}$ & $-0.5$ & 
  $-0.4$ & 
  $-0.3$ & 
  $-0.2$ & 
  $-0.1$ & 
  $0.0$ & 
  $0.1$   \\[3pt] 
\hline

$\mu_{1}$ & $0.100$ & $0.106$ & $0.125$ & $0.075$ & $0.043$ & $0.042$ & $0.021$  \\
$\sigma_{1}$ & $0.027$ & $0.048$ & $0.063$ & $0.049$ & $0.045$ & $0.044$ & $0.038$  \\
$\mu_{2}$ & $0.281$ & $0.275$ & $0.265$ & $0.234$ & $0.174$ & $0.165$ & $0.101$  \\
$\sigma_{2}$ & $0.049$ & $0.042$ & $0.035$ & $0.043$ & $0.045$ & $0.027$ & $0.043$  \\
$R_{21}$ & $7.358$ & $3.592$ & $1.585$ & $1.157$ & $0.833$ & $0.253$ & $0.507$  \\[3pt]

\hline
\hline

\multicolumn{8}{c}{$\mathbf{7\le R < 9 \;\text{\bf kpc}}$} \\
  $\text{[Fe/H]}_{\text{min}}$ & $-0.5$ & 
  $-0.4$ & 
  $-0.3$ & 
  $-0.2$ & 
  $-0.1$ & 
  $0.0$ & 
  $0.1$   \\[3pt] 
\hline

$\mu_{1}$ & $0.111$ & $0.095$ & $0.086$ & $0.057$ & $0.039$ & $0.031$ & $0.038$  \\
$\sigma_{1}$ & $0.038$ & $0.049$ & $0.052$ & $0.046$ & $0.036$ & $0.032$ & $0.031$  \\
$\mu_{2}$ & $0.282$ & $0.273$ & $0.254$ & $0.219$ & $0.171$ & $0.146$ & $0.124$  \\
$\sigma_{2}$ & $0.042$ & $0.036$ & $0.034$ & $0.042$ & $0.049$ & $0.037$ & $0.032$  \\
$R_{21}$ & $0.936$ & $1.074$ & $0.565$ & $0.393$ & $0.338$ & $0.244$ & $0.307$  \\[3pt]

\hline
\hline

\multicolumn{8}{c}{$\mathbf{9\le R < 11 \;\text{\bf kpc}}$} \\
  $\text{[Fe/H]}_{\text{min}}$ & $-0.5$ & 
  $-0.4$ & 
  $-0.3$ & 
  $-0.2$ & 
  $-0.1$ & 
  $0.0$ & 
  $0.1$   \\[3pt] 
\hline

$\mu_{1}$ & $0.092$ & $0.077$ & $0.060$ & $0.034$ & $0.027$ & $0.029$ & $0.035$  \\
$\sigma_{1}$ & $0.033$ & $0.036$ & $0.035$ & $0.031$ & $0.033$ & $0.032$ & $0.027$  \\
$\mu_{2}$ & $0.277$ & $0.265$ & $0.238$ & $0.150$ & $0.145$ & $0.099$ & $0.083$  \\
$\sigma_{2}$ & $0.050$ & $0.045$ & $0.047$ & $0.079$ & $0.060$ & $0.06$ & $0.053$  \\
$R_{21}$ & $0.179$ & $0.166$ & $0.153$ & $0.232$ & $0.193$ & $0.273$ & $0.492$  \\[3pt]

\hline
\hline

\end{tabular}
\caption{Parameters of our best fitting models for $p(\mgfe)$.  Best-fit values of the total dispersions have been converted to {\it intrinsic} values $\sigma_1$, $\sigma_2$ by subtracting in quadrature the median [Mg/Fe] abundance error as reported in APOGEE-DR16.  
}
\end{table*}

\begin{table*}
\centering \label{table2}
\begin{tabular}{c|c|c|c|c|c|c|c}
\hline
\hline
\multicolumn{8}{c}{ {\bf [O/Fe] at $7 \le R < 9\;\text{\bf kpc}$} } \\
   $\text{[Fe/H]}_{\text{min}}$ & $-0.5$ & 
  $-0.4$ & 
  $-0.3$ & 
  $-0.2$ & 
  $-0.1$ & 
  $0.0$ & 
  $0.1$   \\[3pt] 
\hline

$\mu_{1}$ & $0.114$ & $0.095$ & $0.080$ & $0.057$ & $0.040$ & $0.030$ & $0.028$ \\
$\sigma_{1}$ & $0.029$ & $0.030$ & $0.028$ & $0.028$ & $0.024$ & $0.023$ & $0.021$ \\
$\mu_{2}$ & $0.239$ & $0.227$ & $0.208$ & $0.178$ & $0.137$ & $0.113$ & $0.086$ \\
$\sigma_{2}$ & $0.029$ & $0.024$ & $0.023$ & $0.031$ & $0.038$ & $0.030$ & $0.025$ \\[3pt]

\hline
\hline

\multicolumn{8}{c}{ {\bf [Mg/Fe] at $7 \le R < 9\;\text{\bf kpc}$} } \\
  $\text{[Fe/H]}_{\text{min}}$ & $-0.5$ & 
  $-0.4$ & 
  $-0.3$ & 
  $-0.2$ & 
  $-0.1$ & 
  $0.0$ & 
  $0.1$   \\[3pt]
\hline

$\mu_{1}$ & $0.111$ & $0.095$ & $0.086$ & $0.057$ & $0.039$ & $0.031$ & $0.038$  \\
$\sigma_{1}$ & $0.038$ & $0.049$ & $0.052$ & $0.046$ & $0.036$ & $0.032$ & $0.031$  \\
$\mu_{2}$ & $0.282$ & $0.273$ & $0.254$ & $0.219$ & $0.171$ & $0.146$ & $0.124$  \\
$\sigma_{2}$ & $0.042$ & $0.036$ & $0.034$ & $0.042$ & $0.049$ & $0.037$ & $0.032$  \\[3pt]

\hline
\hline

\multicolumn{8}{c}{ {\bf [Si/Fe] at $7 \le R < 9\;\text{\bf kpc}$} } \\
  $\text{[Fe/H]}_{\text{min}}$ & $-0.5$ & 
  $-0.4$ & 
  $-0.3$ & 
  $-0.2$ & 
  $-0.1$ & 
  $0.0$ & 
  $0.1$   \\[3pt]
\hline

$\mu_{1}$ & $0.090$ & $0.075$ & $0.064$ & $0.041$ & $0.024$ & $0.015$ & $0.014$ \\
$\sigma_{1}$ & $0.029$ & $0.031$ & $0.027$ & $0.021$ & $0.020$ & $0.018$ & $0.019$ \\
$\mu_{2}$ & $0.201$ & $0.188$ & $0.170$ & $0.132$ & $0.097$ & $0.075$ & $0.050$ \\
$\sigma_{2}$ & $0.027$ & $0.028$ & $0.026$ & $0.033$ & $0.035$ & $0.025$ & $0.025$ \\[3pt]

\hline
\hline

\multicolumn{8}{c}{ {\bf [S/Fe] at $7 \le R < 9\;\text{\bf kpc}$} } \\
  $\text{[Fe/H]}_{\text{min}}$ & $-0.5$ & 
  $-0.4$ & 
  $-0.3$ & 
  $-0.2$ & 
  $-0.1$ & 
  $0.0$ & 
  $0.1$   \\[3pt]
\hline

$\mu_{1}$ & $0.164$ & $0.133$ & $0.103$ & $0.067$ & $0.038$ & $0.016$ & $0.004$ \\
$\sigma_{1}$ & $0.052$ & $0.057$ & $0.045$ & $0.037$ & $0.028$ & $0.024$ & $0.022$ \\
$\mu_{2}$ & $0.287$ & $0.254$ & $0.219$ & $0.180$ & $0.130$ & $0.099$ & $0.066$ \\
$\sigma_{2}$ & $0.085$ & $0.073$ & $0.059$ & $0.043$ & $0.052$ & $0.062$ & $0.040$ \\[3pt]

\hline
\hline

\multicolumn{8}{c}{ {\bf [Ca/Fe] at $7 \le R < 9\;\text{\bf kpc}$} } \\
  $\text{[Fe/H]}_{\text{min}}$ & $-0.5$ & 
  $-0.4$ & 
  $-0.3$ & 
  $-0.2$ & 
  $-0.1$ & 
  $0.0$ & 
  $0.1$   \\[3pt]
\hline

$\mu_{1}$ & $0.043$ & $0.033$ & $0.028$ & $0.016$ & $0.002$ & $-0.006$ & $-0.010$ \\
$\sigma_{1}$ & $0.021$ & $0.025$ & $0.023$ & $0.024$ & $0.022$ & $0.019$ & $0.023$ \\
$\mu_{2}$ & $0.149$ & $0.130$ & $0.112$ & $0.082$ & $0.049$ & $0.014$ & $-$ \\
$\sigma_{2}$ & $0.031$ & $0.031$ & $0.029$ & $0.030$ & $0.038$ & $0.046$ & $-$ \\[3pt]

\hline
\hline

\end{tabular}
\caption{Parameters of our best fitting models for $p(\ofe)$, $p(\sife)$, $p([{\rm S/Fe}])$, and $p(\cafe)$ at the Solar circle.  Best-fit values of the total dispersions have been converted to {\it intrinsic} values $\sigma_1$, $\sigma_2$ by subtracting in quadrature the median [$\alpha$/Fe] abundance error as reported in APOGEE-DR16. 
}
\end{table*}



\section*{Acknowledgments}
We thank the referee, Matthias Steinmetz, for precious comments and suggestions, which improved the quality and clarity of our work. We acknowledge valuable conversations, some recent and some long-ago, with Brett Andrews, Jonathan Bird, Emily Griffith, James Johnson, Jennifer Johnson, Philipp Kempski, Chiaki Kobayashi, Ted Mackereth, Francesca Matteucci, Ralph Schoenrich, and Emanuele Spitoni. 
This work was supported in part by NSF grant AST-1909841.
FV acknowledges the support of a Fellowship from the Center for Cosmology and AstroParticle Physics at The Ohio State University. DW acknowledges the hospitality of the Institute for Advanced Study and the support of the W.M. Keck Foundation.  AM acknowledges support from the ERC Consolidator Grant funding scheme (project ASTEROCHRONOMETRY, G.A. n. 772293).

In this work we have made use of SDSS-IV APOGEE-2 DR16 data. Funding for the Sloan Digital Sky Survey IV has been provided by the Alfred P. Sloan Foundation, the U.S. Department of Energy Office of Science, and the Participating Institutions. SDSS-IV acknowledges
support and resources from the Center for High-Performance Computing at
the University of Utah. The SDSS web site is \url{www.sdss.org}.

SDSS-IV is managed by the Astrophysical Research Consortium for the 
Participating Institutions of the SDSS Collaboration including the 
Brazilian Participation Group, the Carnegie Institution for Science, 
Carnegie Mellon University, the Chilean Participation Group, the French Participation Group, Harvard-Smithsonian Center for Astrophysics, 
Instituto de Astrof\'isica de Canarias, The Johns Hopkins University, Kavli Institute for the Physics and Mathematics of the Universe (IPMU) / 
University of Tokyo, the Korean Participation Group, Lawrence Berkeley National Laboratory, 
Leibniz Institut f\"ur Astrophysik Potsdam (AIP),  
Max-Planck-Institut f\"ur Astronomie (MPIA Heidelberg), 
Max-Planck-Institut f\"ur Astrophysik (MPA Garching), 
Max-Planck-Institut f\"ur Extraterrestrische Physik (MPE), 
National Astronomical Observatories of China, New Mexico State University, 
New York University, University of Notre Dame, 
Observat\'ario Nacional / MCTI, The Ohio State University, 
Pennsylvania State University, Shanghai Astronomical Observatory, 
United Kingdom Participation Group,
Universidad Nacional Aut\'onoma de M\'exico, University of Arizona, 
University of Colorado Boulder, University of Oxford, University of Portsmouth, 
University of Utah, University of Virginia, University of Washington, University of Wisconsin, 
Vanderbilt University, and Yale University.

\section*{Data availability}
The data underlying this article will be shared on reasonable request to the corresponding author.


\begin{thebibliography}{}

\bibitem[Ahumada et al.(2020)]{ahumada2020} Ahumada, R., Prieto, C.~A., Almeida, A., et al.\ 2020, \apjs, 249, 3

\bibitem[\protect\citeauthoryear{Andrews et al.}{2017}]{andrews2017} Andrews B.~H., Weinberg D.~H., Sch{\"o}nrich R., Johnson J.~A., 2017, ApJ, 835, 224

\bibitem[\protect\citeauthoryear{Belfiore et al.}{2019}]{belfiore2019} Belfiore F., Vincenzo F., Maiolino R., Matteucci F., 2019, MNRAS, 487, 456

\bibitem[\protect\citeauthoryear{Belokurov et al.}{2018}]{belokurov2018} Belokurov V., Erkal D., Evans N.~W., Koposov S.~E., Deason A.~J., 2018, MNRAS, 478, 611

\bibitem[\protect\citeauthoryear{Bensby, Feltzing, \& Lundstr{\"o}m}{2003}]{bensby2003} Bensby T., Feltzing S., Lundstr{\"o}m I., 2003, A\&A, 410, 527

\bibitem[\protect\citeauthoryear{Bertran de Lis et al.}{2016}]{bertran2016} Bertran de Lis S., Allende Prieto C., Majewski S.~R., Schiavon R.~P., Holtzman J.~A., Shetrone M., Carrera R., et al., 2016, A\&A, 590, A74

\bibitem[\protect\citeauthoryear{Boissier \& Prantzos}{1999}]{boissier1999} Boissier S., Prantzos N., 1999, MNRAS, 307, 857

\bibitem[\protect\citeauthoryear{Bovy et al.}{2012}]{bovy2012} Bovy J., Rix H.-W., Liu C., Hogg D.~W., Beers T.~C., Lee Y.~S., 2012, ApJ, 753, 148

\bibitem[Bovy et al.(2016)]{bovy2016} Bovy, J., Rix, H.-W., Schlafly, E.~F., et al.\ 2016, \apj, 823, 30

\bibitem[Bressan et al.(2012)]{bressan2012} Bressan, A., Marigo, P., Girardi, L., et al.\ 2012, \mnras, 427, 127

\bibitem[\protect\citeauthoryear{Buck}{2020}]{buck2020} Buck T., 2020, MNRAS, 491, 5435

\bibitem[Buder et al.(2018)]{buder2018} Buder, S., Asplund, M., Duong, L., et al.\ 2018, \mnras, 478, 4513

\bibitem[\protect\citeauthoryear{Buder et al.}{2020}]{buder2020} Buder S., Sharma S., Kos J., Amarsi A.~M., Nordlander T., Lind K., Martell S.~L., et al., 2020, preprint (arXiv:2011.02505)


\bibitem[\protect\citeauthoryear{Cescutti et al.}{2007}]{cescutti2007} Cescutti G., Matteucci F., Fran{\c{c}}ois P., Chiappini C., 2007, A\&A, 462, 943

\bibitem[\protect\citeauthoryear{Chaplin et al.}{2020}]{chaplin2020} Chaplin W.~J., Serenelli A.~M., Miglio A., Morel T., Mackereth J.~T., Vincenzo F., Kjeldsen H., et al., 2020, NatAs, 4, 382


\bibitem[Chen et al.(2015)]{chen2015} Chen, Y., Bressan, A., Girardi, L., et al.\ 2015, \mnras, 452, 1068

\bibitem[\protect\citeauthoryear{Chiappini, Matteucci, \& Gratton}{1997}]{chiappini1997} Chiappini C., Matteucci F., Gratton R., 1997, ApJ, 477, 765

\bibitem[\protect\citeauthoryear{Clarke et al.}{2019}]{clarke2019} Clarke A.~J., Debattista V.~P., Nidever D.~L., Loebman S.~R., Simons R.~C., Kassin S., Du M., et al., 2019, MNRAS, 484, 3476


\bibitem[\protect\citeauthoryear{Edvardsson et al.}{1993}]{edvardsson1993} Edvardsson B., Andersen J., Gustafsson B., Lambert D.~L., Nissen P.~E., Tomkin J., 1993, A\&A, 500, 391

\bibitem[Foreman-Mackey et al.(2013)]{emcee2013} Foreman-Mackey, D., Hogg, D.~W., Lang, D., et al.\ 2013, \pasp, 125, 306

\bibitem[\protect\citeauthoryear{Fuhrmann}{1998}]{fuhrmann1998} Fuhrmann K., 1998, A\&A, 338, 161

\bibitem[\protect\citeauthoryear{Garc{\'\i}a P{\'e}rez et al.}{2016}]{garcia-perez2016} Garc{\'\i}a P{\'e}rez A.~E., Allende Prieto C., Holtzman J.~A., Shetrone M., M{\'e}sz{\'a}ros S., Bizyaev D., Carrera R., et al., 2016, AJ, 151, 144


\bibitem[\protect\citeauthoryear{Gilmore \& Reid}{1983}]{gilmore1983} Gilmore G., Reid N., 1983, MNRAS, 202, 1025

\bibitem[Girardi(1999)]{girardi1999} Girardi, L.\ 1999, \mnras, 308, 818

\bibitem[Girardi(2016)]{girardi2016} Girardi, L.\ 2016, \araa, 54, 95

\bibitem[\protect\citeauthoryear{Grand et al.}{2018}]{grand2018} Grand R.~J.~J., Bustamante S., G{\'o}mez F.~A., Kawata D., Marinacci F., Pakmor R., Rix H.-W., et al., 2018, MNRAS, 474, 3629

\bibitem[\protect\citeauthoryear{Griffith et al.}{2020}]{griffith2020} Griffith E., Weinberg D.~H., Johnson J.~A., Beaton R., Garc{\'\i}a-Hern{\'a}ndez D.~A., Hasselquist S., Holtzman J., et al., 2020, preprint (arXiv:2009.05063)

\bibitem[\protect\citeauthoryear{Guiglion et al.}{2020}]{guiglion2020} Guiglion G., Matijevi{\v{c}} G., Queiroz A.~B.~A., Valentini M., Steinmetz M., Chiappini C., Grebel E.~K., et al., 2020, A\&A, 644, A168

\bibitem[\protect\citeauthoryear{J{\"o}nsson et al.}{2020}]{jonsson2020} J{\"o}nsson H., Holtzman J.~A., Allende Prieto C., Cunha K., Garc{\'\i}a-Hern{\'a}ndez D.~A., Hasselquist S., Masseron T., et al., 2020, AJ, 160, 120


\bibitem[\protect\citeauthoryear{Hayden et al.}{2015}]{hayden2015} Hayden M.~R., Bovy J., Holtzman J.~A., Nidever D.~L., Bird J.~C., Weinberg D.~H., Andrews B.~H., et al., 2015, ApJ, 808, 132

\bibitem[\protect\citeauthoryear{Haywood et al.}{2013}]{haywood2013} Haywood M., Di Matteo P., Lehnert M.~D., Katz D., G{\'o}mez A., 2013, A\&A, 560, A109

\bibitem[\protect\citeauthoryear{Haywood et al.}{2018}]{haywood2018} Haywood M., Di Matteo P., Lehnert M.~D., Snaith O., Khoperskov S., G{\'o}mez A., 2018, ApJ, 863, 113

\bibitem[\protect\citeauthoryear{Helmi et al.}{2018}]{helmi2018} Helmi A., Babusiaux C., Koppelman H.~H., Massari D., Veljanoski J., Brown A.~G.~A., 2018, Nature, 563, 85

\bibitem[\protect\citeauthoryear{Holtzman et al.}{2015}]{holtzman2015} Holtzman J.~A., Shetrone M., Johnson J.~A., Allende Prieto C., Anders F., Andrews B., Beers T.~C., et al., 2015, AJ, 150, 148

\bibitem[\protect\citeauthoryear{Khoperskov et al.}{2020}]{khoperskov2020} Khoperskov S., Haywood M., Snaith O., Di Matteo P., Lehnert M., Vasiliev E., Naroenkov S., et al., 2020, preprint (arXiv:2006.10195)

\bibitem[Kroupa(2001)]{kroupa2001} Kroupa, P.\ 2001, \mnras, 322, 231

\bibitem[\protect\citeauthoryear{Krumholz \& Ting}{2018}]{krumholzting2018} Krumholz M.~R., Ting Y.-S., 2018, MNRAS, 475, 2236

\bibitem[\protect\citeauthoryear{Lee et al.}{2011}]{lee2011} Lee Y.~S., Beers T.~C., An D., Ivezi{\'c} {\v{Z}}., Just A., Rockosi C.~M., Morrison H.~L., et al., 2011, ApJ, 738, 187

\bibitem[\protect\citeauthoryear{Leung \& Bovy}{2019a}]{leung2019a} Leung H.~W., Bovy J., 2019, MNRAS, 483, 3255


\bibitem[\protect\citeauthoryear{Leung \& Bovy}{2019b}]{leung2019b} Leung H.~W., Bovy J., 2019, MNRAS, 489, 2079

\bibitem[\protect\citeauthoryear{Lian et al.}{2020}]{lian2020} Lian J., Thomas D., Maraston C., Zamora O., Tayar J., Pan K., Tissera P., et al., 2020, MNRAS, 494, 2561

\bibitem[\protect\citeauthoryear{Mackereth et al.}{2017}]{mackereth2017} Mackereth J.~T., Bovy J., Schiavon R.~P., Zasowski G., Cunha K., Frinchaboy P.~M., Garc{\'\i}a Perez A.~E., et al., 2017, MNRAS, 471, 3057

\bibitem[\protect\citeauthoryear{Mackereth et al.}{2019}]{mackereth2018} Mackereth J.~T., Schiavon R.~P., Pfeffer J., Hayes C.~R., Bovy J., Anguiano B., Allende Prieto C., et al., 2019, MNRAS, 482, 3426

\bibitem[\protect\citeauthoryear{Majewski et al.}{2017}]{majewski2017} Majewski S.~R., Schiavon R.~P., Frinchaboy P.~M., Allende Prieto C., Barkhouser R., Bizyaev D., Blank B., et al., 2017, AJ, 154, 94


\bibitem[Matteucci \& Greggio(1986)]{matteucci1986} Matteucci, F. \& Greggio, L.\ 1986, \aap, 154, 279

\bibitem[\protect\citeauthoryear{Matteucci \& Francois}{1989}]{matteucci1989} Matteucci F., Francois P., 1989, MNRAS, 239, 885

\bibitem[\protect\citeauthoryear{Matteucci}{2012}]{matteucci2012} Matteucci F., 2012, Chemical Evolution of Galaxies. Astrophysics and Space Science Library, Springer-Verlag, Berlin

\bibitem[\protect\citeauthoryear{McWilliam}{1997}]{mcwilliam1997} McWilliam A., 1997, ARA\&A, 35, 503

\bibitem[Miglio et al.(2020)]{miglio2020} Miglio, A., Chiappini, C., Mackereth, T., et al.\ 2020, preprint (arXiv:2004.14806)

\bibitem[\protect\citeauthoryear{Minchev, Chiappini, \& Martig}{2013}]{minchev2013} Minchev I., Chiappini C., Martig M., 2013, A\&A, 558, A9 

\bibitem[\protect\citeauthoryear{Montalb{\'a}n et al.}{2020}]{montalban2020} Montalb{\'a}n J., Mackereth J.~T., Miglio A., Vincenzo F., Chiappini C., Buldgen G., Mosser B., et al., 2020, preprint (arXiv:2006.01783)

\bibitem[\protect\citeauthoryear{Nandakumar et al.}{2020}]{nandakumar2020} Nandakumar G., Hayden M.~R., Sharma S., Buder S., Asplund M., Bland-Hawthorn J., De Silva G.~M., et al., 2020, preprint (arXiv:2011.02783)

\bibitem[\protect\citeauthoryear{Nidever et al.}{2014}]{nidever2014} Nidever D.~L., Bovy J., Bird J.~C., Andrews B.~H., Hayden M., Holtzman J., Majewski S.~R., et al., 2014, ApJ, 796, 38

\bibitem[\protect\citeauthoryear{Noguchi}{2018}]{noguchi2018} Noguchi M., 2018, Nature, 559, 585

\bibitem[\protect\citeauthoryear{Palla et al.}{2020}]{palla2020} Palla M., Matteucci F., Spitoni E., Vincenzo F., Grisoni V., 2020, MNRAS, 498, 1710

\bibitem[\protect\citeauthoryear{Pontzen et al.}{2017}]{pontzen2017} Pontzen A., Tremmel M., Roth N., Peiris H.~V., Saintonge A., Volonteri M., Quinn T., et al., 2017, MNRAS, 465, 547

\bibitem[\protect\citeauthoryear{Sch{\"o}nrich \& Binney}{2009}]{schoenrich2009} Sch{\"o}nrich R., Binney J., 2009, MNRAS, 396, 203

\bibitem[\protect\citeauthoryear{Sharma, Hayden, \& Bland-Hawthorn}{2020}]{sharma2020} Sharma S., Hayden M.~R., Bland-Hawthorn J., 2020, preprint (arXiv:2005.03646)

\bibitem[\protect\citeauthoryear{Spitoni et al.}{2019}]{spitoni2019} Spitoni E., Silva Aguirre V., Matteucci F., Calura F., Grisoni V., 2019, A\&A, 623, A60

\bibitem[Tang et al.(2014)]{tang2014} Tang, J., Bressan, A., Rosenfield, P., et al.\ 2014, \mnras, 445, 4287

\bibitem[\protect\citeauthoryear{Tinsley}{1980}]{tinsley1980} Tinsley B.~M., 1980, FCPh, 5, 287

\bibitem[Weinberg et al.(2017)]{weinberg2017} Weinberg, D.~H., Andrews, B.~H., \& Freudenburg, J.\ 2017, \apj, 837, 183

\bibitem[\protect\citeauthoryear{Weinberg et al.}{2019}]{weinberg2019} Weinberg D.~H., Holtzman J.~A., Hasselquist S., Bird J.~C., Johnson J.~A., Shetrone M., Sobeck J., et al., 2019, ApJ, 874, 102


\bibitem[\protect\citeauthoryear{Vincenzo et al.}{2019}]{vincenzo2019} Vincenzo F., Spitoni E., Calura F., Matteucci F., Silva Aguirre V., Miglio A., Cescutti G., 2019, MNRAS, 487, L47

\bibitem[Vincenzo \& Kobayashi(2020)]{vincenzo2020} Vincenzo, F. \& Kobayashi, C.\ 2020, \mnras, 496, 80

\bibitem[\protect\citeauthoryear{Wheeler et al.}{2020}]{wheeler2020} Wheeler A., Ness M., Buder S., Bland-Hawthorn J., Silva G.~D., Hayden M., Kos J., et al., 2020, ApJ, 898, 58

\bibitem[\protect\citeauthoryear{Xiang et al.}{2019}]{Xiang2019} Xiang M., Ting Y.-S., Rix H.-W., Sandford N., Buder S., Lind K., Liu X.-W., et al., 2019, ApJS, 245, 34

\bibitem[\protect\citeauthoryear{Zasowski et al.}{2013}]{zasowski2013} Zasowski G., Johnson J.~A., Frinchaboy P.~M., Majewski S.~R., Nidever D.~L., Rocha Pinto H.~J., Girardi L., et al., 2013, AJ, 146, 81

\bibitem[\protect\citeauthoryear{Zasowski et al.}{2017}]{zasowski2017} Zasowski G., Cohen R.~E., Chojnowski S.~D., Santana F., Oelkers R.~J., Andrews B., Beaton R.~L., et al., 2017, AJ, 154, 198




\end{thebibliography}
\end{document}